\documentclass[preprint,12pt]{elsarticle}

\usepackage{amssymb}
\usepackage{amsmath,amssymb,amsfonts}
\usepackage[shortcuts]{extdash}
\usepackage{algorithm}
\usepackage{algpseudocode}
\usepackage[normalem]{ulem}
\usepackage{multicol}
\usepackage{multirow}
\usepackage{ulem}
\usepackage{xcolor}

\newenvironment{ldescription}[1]
  {\begin{list}{}%
   {\renewcommand\makelabel[1]{##1\hfill}%
   \settowidth\labelwidth{\makelabel{#1}}%
   \setlength\leftmargin{\labelwidth}
   \addtolength\leftmargin{\labelsep}}}
  {\end{list}}

\journal{Elsevier}

\begin{document}

\begin{frontmatter}



\title{Impact of Phase Selection on Accuracy and Scalability in Calculating Distributed Energy Resources Hosting Capacity}


\author[a]{Tomislav Antić}
\author[b]{Andrew Keane}
\author[a]{Tomislav Capuder}

\affiliation[a]{organization={University of Zagreb Faculty of Electrical Engineering and Computing},
            addressline={Unska 3}, 
            city={Zagreb},
            country={Croatia}}

\affiliation[b]{organization={University College Dublin, UCD Energy Institute},
            addressline={Belfield}, 
            city={Dublin 4},
            country={Ireland}}

\begin{abstract}
Hosting capacity (HC) and dynamic operating envelopes (DOEs), defined as dynamic, time-varying HC, are calculated using three-phase optimal power flow (OPF) formulations. Due to the computational complexity of such optimisation problems, HC and DOE are often calculated by introducing certain assumptions and approximations, including the linearised OPF formulation, which we implement in the Python-based tool ppOPF. Furthermore, we investigate how assumptions of the distributed energy resource (DER) connection phase impact the objective function value and computational time in calculating HC and DOE in distribution networks of different sizes. The results are not unambiguous and show that it is not possible to determine the optimal connection phase without introducing binary variables since, no matter the case study, the highest objective function values are calculated with mixed integer OPF formulations. The difference is especially visible in a real-world low-voltage network in which the difference between different scenarios is up to 14 MW in a single day. However, binary variables make the problem computationally complex and increase computational time to several hours in the DOE calculation, even when the optimality gap different from zero is set.
\end{abstract}


\begin{highlights}
\item Optimal power flow-based hosting capacity calculation is computationally complex
\item We investigate the impact of different assumptions on the accuracy and scalability
\item Simulations are performed by ppOPF, tool developed as the pandapower extension
\item The Python-based tool ppOPF is extended by implementing the linearised formulation
\item Same investigation is performed for dynamic operating envelopes for the first time
\end{highlights}


\end{frontmatter}


\section*{Nomenclature}

\subsection*{Sets and Indices}
\begin{ldescription}{$xxxxxx$}
    \item[$N$] Set of all nodes
    \item[$P$] Set of all phases of the node
    \item[$T$] Set of all time intervals
    \item[$ij$] Set of nodes connected by a branch, $i, j \in N$
    \item[$n$] Observed node, $n \in N$
    \item[$p$] Phase of the node, $p \in P$
    \item[$t$] $t \in T$
\end{ldescription}

\subsection*{Parameters}
\begin{ldescription}{$xxxxxxxxxxxxxxxx$}
    \item[$P_{n,p,t}^{load}$] Load active power
    \item[$Q_{n,p,t}^{load}$] Load reactive power
    \item[$R_{ij,pq}$] Branch $ij$ resistance between phases $pq$
    \item[$X_{ij,pq}$] Branch $ij$ reactance between phases $pq$
    \item[$U^{min/max}$] Minimum/maximum voltage magnitude, 0.9 p.u./1.1 p.u.
    \item[$VUF^{max}$] Maximum voltage unbalance factor, 2\%
    \item[$I_{ij}^{max}$] Maximum current flow in element $ij$
\end{ldescription}
\subsection*{Variables}
\begin{ldescription}{$xxxxxxxxx$}
    \item[$U_{n,p,t}^{re/im}$] Real/imaginary part of voltage at node $n$, phase $p$, and time period $t$
    \item [$W_{n,t}$] Square voltage magnitude at node $n$ and time period $t$
    \item[$I_{ij,p,t}^{re/im}$] Real/imaginary part of current of branch $ij$, phase $p$, and at time period $t$
    \item[$P_{ij,p,t}$] Active power flow in branch $ij$, phase $p$, and at time period $t$
    \item[$Q_{ij,p,t}$] Reactive power flow in branch $ij$, phase $p$, and at time period $t$
    \item[$\left(I_{n,p,t}^{load}\right)^{re/im}$] Real/imaginary part of load's current connected to the node $n$, phase $p$, at time period $t$
    \item[$\left(I_{n,p,t}^{gen}\right)^{re/im}$] Real/imaginary part of generator's current connected to the node $n$, phase $p$, at time period $t$
    \item[$\left(I_{n,p,t}^{DER}\right)^{re/im}$] Real/imaginary part of DER's current connected to the node $n$, phase $p$, at time period $t$
    \item[$P_{n,p,t}^{gen/DER}$] Active power of load/DER connected to the node $n$, phase $p$, at time period $t$
    \item[$Q_{n,p,t}^{gen/DER}$] Reactive power of of load/DER connected to the node $n$, phase $p$, at time period $t$
    \item[$x^{DER}_{n,p,t}$] Binary variable defining if DER is connected to phase $p$ at node $n$ in time period $t$
\end{ldescription}

\section{Introduction} \label{sec:intro}
\subsection{Motivation}
In line with the European policies and green energy transition strategies, the share of renewable energy sources is continuously increasing \cite{IRENA2024}. Due to the decrease in prices and different governments' stimulation, the installed renewable, and especially photovoltaic (PV), capacity is also growing \cite{IEA2022}. Besides the number and capacity of PVs, more residential electricity end-users decide to invest in the electrification of their heating and transport technologies, leading to a growth in the share of heat pumps (HPs) and electric heaters \cite{IEA2023} and electric vehicles (EVs), and consequentially, EV charging stations \cite{IEA2023EV}. Despite the environmental and potential financial benefits of distributed energy resources (DERs) integration, it is often uncoordinated and does not follow proper technical analyses, violating the network's technical limitations and creating problems in the planning and operation of distribution system operators (DSOs) \cite{ANTIC2022100926}. 

Even though mitigating technical issues is possible by strict connection limits defined in national distribution system grid codes \cite{grid_code}, such definitions are often conservative and unnecessarily limit the integration of DERs, especially in the case of a single-phase connection. To move aside from traditional approaches and conservative limitations, the concept of hosting capacity (HC), in which network limitations directly impact the installed power of DERs, is being recognized by DSOs as a method that allows the increase in the share of DERs but still does not endanger the safe network's operation \cite{10415382}.

\subsection{Literature Review}
HC is being recognized as a valid approach to assess the potential of DER integration in distribution networks. The benefit of the approach is that it can be used to calculate both export limits, e.g., the installed capacity of PVs \cite{YAO2022119681}, import limits, e.g., installed power of EV charging stations or HPs \cite{EDMUNDS2021117093}, or their combination \cite{FACHRIZAL2021100445}. Due to the simplicity and the installation cost, single-phase inverters through which DERs are connected to a network are preferable from the end-users' standpoint. Without the proper investigation of the technical conditions after the single-phase connection, significant voltage unbalance values are possible \cite{en14010117}. Therefore, the voltage unbalance constraint needs to be included when calculating the single-phase HC. The analysis of a realistically modelled LV network demonstrates that the large voltage unbalance limits EV hosting capacity \cite{10102721}. The same results are shown for the case of PVs, where rebalancing ensures the adoption of additional PVs at the expense of the voltage unbalance. Toghranegar et al. minimise the voltage unbalance value and consequentially increase HC of DERs (EVs, HPs, and PVs) by applying a three-step approach based on the heuristic optimisation \cite{TOGHRANEGAR2022104243}. However, most of the methods in the published papers rely on stochasticity and randomization techniques applied in determining the DERs connection phase. As shown in our previous work \cite{10257450}, the randomly selected connection phase in the HC calculation does not guarantee a theoretical maximum and leads to a suboptimal solution. In the analysis presented in the mentioned paper \cite{10257450}, our focus was on calculating the export limits and comparison of PV generation values using different approaches, from which only one guaranteed reaching the theoretical maximum and optimality of the solution. In this study, we extend the analysis to calculating import limits and develop several more approaches to investigate the accuracy of the HC calculation.

The reason for not accurately modelling the DERs connection phase in the HC calculation is the complexity of the calculation. HC is often assessed using optimal power flow (OPF) formulations. OPF formulation is nonlinear and nonconvex, making it computationally complex to solve \cite{GETH2020106558}. Combining the complexity of OPF models with specific characteristics of three-phase distribution networks increases the computational burden when solving the optimisation problem. Therefore, OPF formulations are often linearised.
Gan and Low generalise the \textit{Simplified Distflow Equations} from single-phase networks to multiphase networks and show that the voltages are within 0.0016 per unit and the power flows are within 5.3\% of their true values for all test networks \cite{7038399}. Another modification of the DistFlow formulation is proposed in \cite{7741261,7244261}, in which the authors extend the single-phase \textit{LinDistFlow} formulation to an unbalanced case, creating \textit{Lin3DisfFlow} and further relaxing the complexity of the problem. Claeys et al. compare the two approaches and show that the relaxation introduced in \textit{Lin3DisfFlow} does not significantly affect the accuracy when compared to \textit{Unbalanced Simplified Distflow}, but it is less accurate in comparison to the results calculated by the exact, nonlinear three-phase OPF formulation \cite{9639999}. Another, more detailed investigation of the loss of accuracy is presented in \cite{VANIN2020106699}, in which the authors show that using \textit{Unbalanced Simplified Distflow} and SOCP formulations leads to the loss of accuracy but decreases the computational time needed to solve the optimisation problem, especially in test feeders with the higher number of nodes. Moreover, Vanin et al. increase the complexity of the problem by considering the need to decrease electricity demand to enable EV charging without violating voltage limitations, requiring the introduction of binary variables into the optimisation problem \cite{VANIN2020106699}. Introducing binary variables emphasises the increase in the computational time when using the exact mixed integer nonlinear programming (MINLP) formulation in comparison to mixed integer \textit{Unbalanced Simplified Distflow} and mixed integer SOCP. However, the authors did not compare the accuracy and computational time of the same optimisation problem with and without binary variables. Therefore, we propose a method for removing binary variables from the single-phase HC calculation, and we test the accuracy and scalability of MINLP, NLP, MILP, and LP OPF formulations. Furthermore, to the best of the authors' knowledge, we present a first detailed investigation of applying the pyomo optimisation framework and Python programming language in optimisation problems, more specifically, OPF formulations in active distribution networks.

Using direct optimisation approaches, i.e., applying OPF formulations, in calculating HC requires the development of tools with the successfully implemented mathematical model, enabling solving the given optimisation problem. However, there are not many such tools in the literature, especially in the case of the tools applied in three-phase distribution networks. 
There are many analytical tools based on open-source technologies applied in basic and optimal power flow calculations for balanced, single-phase power networks. Matlab-based tool MATPOWER provides a high-level set of power flow, optimal power flow (OPF), and other tools applied in power system analyses \cite{5491276}. 
OpenDSS is the open-source distribution networks simulation tool developed by the Electric Power Research Institute \cite{openDSS}. It supports the integration of DERs in distribution networks by enabling complex analyses, including steady-state power flow calculations. 
Pandapower is a Python-based open source power system analysis tool aimed at automating static and quasi-static analysis and optimisation of balanced and unbalanced power systems \cite{8344496}. 
Recently, there have been multiple extensions of the above-mentioned tools, making them applicable in LV network analyses. Network modelling in OpenDSS is extended with a nonlinear, nonconvex three-phase OPF formulation in the tool Open-DSOPF \cite{9282125}, while existing pandapower functionalities are used in ppOPF, the Python-based tool with implemented current-voltage and power-voltage three-phase OPF formulations \cite{10252870}. However, both of the mentioned tools do not have implemented relaxed or linearised formulations, making the tools questionably applicable in more complex distribution network analyses. 
The open-source framework with the highest number of OPF formulations is PowerModelDistribution.jl, developed as a a Julia package by Fobes et al. \cite{FOBES2020106664}. The framework contains nonlinear, linear and relaxed unbalanced OPF formulations, and it allows DER modelling for distribution networks.
In this work, we further extend the functionalities of ppOPF by implementing the linearised three-phase OPF formulation. Furthermore, we test the newly developed formulation in calculating (dynamic) HC in an LV network. Also, the accuracy of previously integrated nonlinear formulations has been increased by changing the source code. Finally, the improvements and the newly developed linearised formulation are made publicly available \cite{ppOPF}.  

Even though it enables higher integration of DERs, the HC calculation has several shortcomings. The biggest issue with the HC assessment is the definition of the worst-case initial electricity demand scenario, i.e., in the case of calculating export limits, minimum electricity demand is observed, and in the case of import limits, maximum demand is defined.
Therefore, the general focus in research has been put on the concept of dynamic operating envelopes (DOEs), whose calculation relies on day-ahead or real-time demand estimates. By managing DERs, DOE implementation allows an even higher share of DERs, moving away from calculating static export and import limits and introducing dynamic limits while maintaining the network within operational and technical limitations \cite{9715663}. 
It is important to mention that DOEs can also be related to PQ flexibility regions, control of DERs and their contribution to the improvement of technical network conditions. However, in previously mentioned works and this study, DOEs are defined as the dynamic HC or time-varying import and export limits. That way, the assessment of the DER integration is not based on the worst-case scenario as in the HC concept but relies on day-ahead or near real-time electricity demand estimates, creating the space for additional DER power in the network.
Even though the linearisation technique has already been applied in calculating DOE \cite{9816082}, as well as a nonlinear and nonconvex formulation \cite{10202795}, none of the published works in the field of DOE calculation focus on investigating the potential loss of accuracy and scalability issues by approximating the OPF formulation. Therefore, the focus of this paper is analysing the benefits of using the (MI)NLP formulation and how some assumptions impact the value of the objective function and computational time.

\subsection{Contributions}
The methods of calculating HC have been properly investigated in the literature, and the focus of the lately published papers is on the techniques for increasing the DER export and import limits. One of the ways to enable further DER integration is the DOE assessment. However, both in the case of HC and DOEs, research lacks the proper investigation of the impact of DER phase connection on the objective function value. Even though it ensures the optimality of the solution, it requires the introduction of binary variables, uplifting the formulation to MINLP, which increases the computational time needed to solve the optimisation problem. Therefore, we present several approaches to avoid binary variables and compare the results of MINLP to NLP and also to MILP and LP formulations. Furthermore, we upgrade pp OPF \cite{ppOPF}, the Python-based tool developed as the extension of pandapower. We improve the accuracy of existing, already implemented nonlinear formulations and develop the linearised formulation as a part of the tool. The tool is made open-source and publicly available, allowing others to reproduce the presented analysis but also to further investigate similar problems that can be solved by applying the three-phase OPF formulation.

After detailed literature review and identified research gaps, we propose following contributions:
\begin{itemize}
    \item The mathematical formulation for determining the optimal phase connection in the DOE and HC calculation based on the three-phase OPF model.
    \item Investigation of the impact of using (MI)NLP and (MI)LP OPF formulations on the accuracy and scalability in calculating HC and DOE values.
    \item Extension of the Python-based open-source solution pp OPF, by increasing the accuracy of current\=/voltage and power-voltage NLP formulations and the novel implementation of linearised three-phase OPF model.
\end{itemize}

The rest of the paper is organised as follows: detailed mathematical formulation of all used OPF models and explanation of selecting a DER connection phase is presented in Section \ref{sec:methodology}, test networks, case studies, and scenarios are described in Section \ref{sec:cs}, results are shown in Section \ref{sec:results}, while the conclusions are given in Section \ref{sec:conclusion}.

\section{Methodology} \label{sec:methodology}
In this paper, we calculate HC and DOE by applying different three-phase OPF formulations implemented in the Python-based open-source tool ppOPF. One of the goals of the study is to assess how different OPF formulations impact the value of the objective function and the computational time needed to solve the optimisation problem. Three OPF formulations are implemented and tested, two of which are nonlinear and nonconvex, and one is linearised.
The mathematical formulation and open-source implementation of the nonlinear current-voltage formulation is based on the model presented in \cite{9133699} and the applied linearisation technique is demonstrated in \cite{7741261}. In this section, only the fundamental set of equations relevant to understanding the difference between the formulations is presented.

\subsection{Current-voltage OPF formulation}
The real and imaginary part of the voltage drop of each line connecting nodes $i$ and $j$ is shown with \eqref{eq:voltage_drop_re}-\eqref{eq:voltage_drop_im}.

\begin{equation}
    U_{j,p,t}^{re} = U_{i,p,t}^{re} - \sum_{q \in \{a,b,c\}}R_{l,pq}\cdot I_{l,ij,q,t}^{re} + \sum_{q \in \{a,b,c\}}X_{l,pq}\cdot I_{l,ij,q,t}^{im}   
\label{eq:voltage_drop_re}
\end{equation}

\begin{equation}
    U_{j,p,t}^{im} = U_{i,p,t}^{im} - \sum_{q \in \{a,b,c\}}R_{l,pq}\cdot I_{l,ij,q,t}^{im} - \sum_{q \in \{a,b,c\}}X_{l,pq}\cdot I_{l,ij,q,t}^{re}  
\label{eq:voltage_drop_im}
\end{equation}

Active and reactive power flow in the branch $ij$ are calculated from the correlation between voltage and current values, as shown with \eqref{eq:line_act_power_calc_phase}-\eqref{eq:line_react_power_calc_phase}.

\begin{equation}
    P_{ij,p,t} = U_{i,p,t}^{re}\cdot I_{ij,p,t}^{re} + U_{i,p,t}^{im}\cdot I_{ij,p,t}^{im} 
    \label{eq:line_act_power_calc_phase}
\end{equation}
\begin{equation}
    Q_{ij,p,t} = U_{i,p,t}^{im}\cdot I_{ij,p,t}^{re} - U_{i,p,t}^{re}\cdot I_{ij,p,t}^{im} 
    \label{eq:line_react_power_calc_phase}
\end{equation}

As will be seen later, the objective function of the optimisation problem is maximising static and dynamic DER connection limits in terms of power. Furthermore, input values of existing loads and generators in a network are given through active and reactive powers. However, the formulation is based on currents and voltages, and therefore, the correlation between node voltage, elements current, and their power is defined \eqref{eq:p_element}-\eqref{eq:q_element}.

\begin{equation}
    P_{n,p,t}^{load/gen/DER} = U_{n,p,t}^{re} \cdot \left(I_{n,p,t}^{load/gen/DER}\right)^{re} + U_{n,p,t}^{im} \cdot \left(I_{n,p,t}^{load/gen/DER}\right)^{im}
\label{eq:p_element}
\end{equation}
\begin{equation}
   Q_{n,p,t}^{load/gen/DER} = U_{n,p,t}^{im} \cdot \left(I_{n,p,t}^{load/gen/DER}\right)^{re} - U_{n,p,t}^{re} \cdot \left(I_{n,p,t}^{load/gen/DER}\right)^{im}
\label{eq:q_element}
\end{equation}

Variable $P_{n,p,t}^{DER}$ is defined as unbounded in the presented formulation. Even though DSOs often limit DERs connection power, defined limits are often conservative, resulting in the total installed power being significantly lower than the network can accommodate. Therefore, we abandon this constraint and assess the HC value based only on network conditions. 

Kirchoff's Current Law is ensured with (\ref{eq:kirchoff_current}), and it is valid for both the real and imaginary parts of the current. The sign in front of the DER's current can change depending on whether a DER is a generator or a consumption unit.
\begin{equation}
    \left(I_{i,p,t }^{load}\right)^{re/im} - \left(I_{i,p,t }^{gen}\right)^{re/im} \pm \left (I_{i,p,t }^{DER}\right)^{re/im} - I_{l,h \rightarrow i,p,t}^{re/im} + I_{l,i \rightarrow j,p,t}^{re/im} = 0
    \label{eq:kirchoff_current}
\end{equation}

In calculating DER HC and DOE, it is important to bound the values of technical indicators limiting the total installed DER power. Limitations are put on the values of voltage magnitude, branch's current flow and voltage unbalance factor (VUF). Since the DER connection power is unbounded in the formulation, without defining the threshold values for relevant technical indicators, DSOs would face the risk of endangering the safe and reliable operation of a network. Therefore, it is important to introduce constraints defined by \eqref{eq:node_volt_limit}-\eqref{eq:max_current} and ensure that the network is operated within allowed limitations. According to the Croatian distribution system grid code \cite{grid_code} and relevant standards EN 50160 \cite{en50160} and IEC 61000-2-2 \cite{iec61000_2_2} voltage magnitude needs to be between minimum and maximum defined voltage, i.e., 90\% and 110\% of the nominal voltage value \eqref{eq:node_volt_limit}. In the same set of standards, the threshold VUF value of 2\% is also defined \eqref{eq:vuf}. Also, the standards define that VUF is calculated as the ratio between negative and positive sequence voltage values. In the presented OPF formulation, real and imaginary parts of voltage magnitude are calculated for each of the three phases and not for the negative, positive and zero sequence system values. However, by applying the common transformation technique, it is relatively easy to adapt \eqref{eq:vuf} and use phase values in the VUF calculation. The current flow is defined by each element's ampacity, as defined with \eqref{eq:max_current}.

\begin{equation}
    (U^{min})^2 \leq (U_{n,p,t}^{re})^2 + (U_{n,p,t}^{im})^2 \leq (U^{max})^2
    \label{eq:node_volt_limit}
\end{equation}

\begin{equation}
    \frac{|U_{n,t}^{negative}|^2}{|U_{n,t}^{positive}|^2} \leq (VUF_n^{max})^2
    \label{eq:vuf}
\end{equation}

\begin{equation}
    \left(I_{l,ij,p,t}^{re}\right)^2 + \left(I_{l,ij,p,t}^{im}\right)^2 \leq \left(I_{l,ij}^{max}\right)^2
    \label{eq:max_current}
\end{equation}

\subsection{Linearised OPF formulation}
Linearised formulation approximates the constraints in which two variables are multiplied. Since the linearisation of the current-voltage formulation is more complex, we implemented the linearised power-voltage formulation. The presented formulation is commonly known as \textit{LinDist3Flow}. In the first step, the product of multiplying voltage vectors is replaced with the variable $W$ that will be used in the rest of the formulation \eqref{eq:square_vol}.

\begin{equation}
    \left[U_{n,t}\right] \cdot \left[U_{n,t}\right]^H = \left[W_{n,t}\right]
    \label{eq:square_vol}
\end{equation}

Eq. \eqref{eq:lin_volt_drop} shows the calculation of the voltage drop between nodes $i$ and $j$. In this formulation, the traditional impedance matrix is modified by applying the transformation described in \cite{7038399,7741261,7244261}.

\begin{equation}
\begin{gathered}
    \left[W_{j,t}\right] = \left[W_{i,t}\right] - \\
    \begin{bmatrix}
        -2R_{ij,aa} & R_{ij,ab} - \sqrt{3}X_{ij,ab} & R_{ij,ac} + \sqrt{3}X_{ij,ac} \\
        R_{ij,ba} + \sqrt{3}X_{ij,ba} & -2R_{ij,bb} & R_{ij,bc} - \sqrt{3}X_{ij,bc} \\
        R_{ij,ca} - \sqrt{3}X_{ij,ca} & R_{ij,cb} + \sqrt{3}X_{ij,cb} & -2R_{ij,cc}
        \end{bmatrix} \cdot 
        \begin{bmatrix}
            P_{ij,a,t} \\
            P_{ij,b,t} \\
            P_{ij,c,t}
        \end{bmatrix} - \\
    \begin{bmatrix}
        -2X_{ij,aa} & R_{ij,ab} + \sqrt{3}R_{ij,ab} & X_{ij,ac} - \sqrt{3}R_{ij,ac} \\
        X_{ij,ba} - \sqrt{3}R_{ij,ba} & -2X_{ij,bb} & X_{ij,bc} + \sqrt{3}R_{ij,bc} \\
        X_{ij,ca} + \sqrt{3}R_{ij,ca} & X_{ij,cb} - \sqrt{3}R_{ij,cb} & -2X_{ij,cc}
        \end{bmatrix} \cdot 
        \begin{bmatrix}
            Q_{ij,a,t} \\
            Q_{ij,b,t} \\
            Q_{ij,c,t}
        \end{bmatrix}
\end{gathered}
\label{eq:lin_volt_drop}
\end{equation}

Kirchoff's Current Law is expressed with active and reactive power values \eqref{eq:kirchoff_active}-\eqref{eq:kirchoff_reactive}. However, the equality defined in the constraint remains the same as in the current-voltage formulation.

\begin{equation}
    P_{i,p,t}^{load} - P_{i,p,t}^{gen} \pm P_{i,p,t}^{DER} - P_{h \rightarrow i,p,t} + P_{i \rightarrow j,p,t}
    \label{eq:kirchoff_active}
\end{equation}

\begin{equation}
    Q_{i,p,t}^{load} - Q_{i,p,t}^{gen} \pm Q_{i,p,t}^{DER} - Q_{h \rightarrow i,p,t} + Q_{i \rightarrow j,p,t}
    \label{eq:kirchoff_reactive}
\end{equation}

Voltage magnitude is constrained by \eqref{eq:w_volt}, in which minimum and maximum voltage values remain within 90\% and 110\% of the nominal voltage value.

\begin{equation}
    (U^{min})^2 \leq W_{n,p,t} \leq (U^{max})^2
    \label{eq:w_volt}
\end{equation}

As mentioned before, linearisation is an approximation technique, meaning that some physical laws and network limitations are neglected. Such a case is with branch flow and voltage unbalance limitations. Even though some approximation techniques define these limitations, the \textit{LindDist3Flow} formulation does not consider them. If voltage magnitude limitation is the constraining one, made assumptions are not a problem for the algorithm accuracy. However, if branch flow or voltage unbalance are thresholds met first, calculated HC or DOE values would potentially exceed ones calculated by the exact nonlinear formulation. Therefore, the emphasis of the accuracy investigation will be on the (MI)LP formulation and the results obtained by using it.

\subsection{Determining a DER connection}
HC is calculated when the initial electricity demand in a network is minimum or maximum, depending on whether the export or import limits are calculated. We define the initial electricity demand in the following way: for each end-user connected at the node $n$, and for each phase $p$, we find the time period defined by the minimum or maximum electricity demand \eqref{eq:p_min_max_user_phase} and after that, minimum or maximum phase demands are summed for each end-user in a network \eqref{eq:p_min_max_user}. That way, we ensure that HC is calculated for the edge scenario and that network constraints will not be violated in the real-time operation. In the DOE calculation, it is not necessary to determine the maximum or minimum electricity demand since, in this study, DOE is assessed for every time period based on the day-ahead estimates.

\begin{equation}
    \left(P_{n,p}^{load}\right)^{min/max} = min/max \ P_{n,p,t}^{load}, \ \forall n \in N, \ \forall p \in P 
    \label{eq:p_min_max_user_phase}
\end{equation}

\begin{equation}
    \left(P_{n}^{load}\right)^{min/max} = \left(P_{n,p}^{load}\right)^{min/max}, \ \forall n \in N
    \label{eq:p_min_max_user}
\end{equation}

After determining the initial electricity demand for each end-user for the HC calculation, it is necessary to select the phase to which DERs will be connected. In the case of a single-phase connected end-user, the DER connection phase is the same as the one to which end-users are connected. In some other cases, end-users can be three-phase connected. In that case, determining the connection phase of DERs is relatively intuitive since they can be connected to each of the three phases. However, mathematically, determining the optimal connection phase is not easy to understand. A randomly selected connection phase can lead to quicker reaching network limitations and threshold values of constraining technical quantities. Therefore, finding an optimal connection phase can be important when maximising the share of DERs in an LV network.

To ensure the optimality of the solution and calculate the theoretically maximum objective function values in the single-phase DER HC and DOE calculation. In the current-voltage formulation, DER current variables in \eqref{eq:p_element}-\eqref{eq:q_element} and \eqref{eq:kirchoff_current} are replaced by the formulation defined in \eqref{eq:binary_i_der}, while in the linearised formulation, we replace DER active and reactive power in the Kirchoff's Current Law defined in \eqref{eq:kirchoff_active}-\eqref{eq:kirchoff_reactive} by the modified formulation presented in \eqref{eq:binary_p_der}-\eqref{eq:binary_q_der}. It needs to be stated that in the model, DERs are not able to provide Volt/Var regulation, i.e., their reactive power is set to zero. Therefore, \eqref{eq:binary_q_der} is only presented from a theoretical standpoint and is not implemented in the model used in analyses.
To ensure a single-phase connection, \eqref{eq:pv_binaries_sum} defines the binary variable at only one phase can have a value equal to 1. $x_{n,p,t}^{DER}$ changes over time, i.e., it is not required for a DER to be connected to the same phase in each period. That way, it is possible to calculate the theoretically maximal export and import limits in an observed network. It needs to be stated that changing the connection phase does not significantly impact the voltage unbalance since its value is bounded and cannot be larger than a defined threshold value. However, it can lead to an increase in the objective function value when compared to the fixed connection phase.

\begin{equation}
    \left(\left(I_{n,p,t}^{DER}\right)^{re/im}\right)' = x_{n,p,t}^{DER}\cdot \left(I^{DER}_{n,p,t}\right)^{re/im}
    \label{eq:binary_i_der}
\end{equation}

\begin{equation}
    \left(P_{n,p,t}^{DER}\right)' = x_{n,p,t}^{DER}\cdot  P_{n,p,t}^{DER}
    \label{eq:binary_p_der}
\end{equation}

\begin{equation}
    \left(Q_{n,p,t}^{DER}\right)' = x_{n,p,t}^{DER}\cdot  Q_{n,p,t}^{DER}
    \label{eq:binary_q_der}
\end{equation}

\begin{equation}
    \sum_{p \in \{a,b,c\}} x^{DER}_{n,p,t} \leq 1, \ \forall n \in N, \ \forall t \in T
    \label{eq:pv_binaries_sum}
\end{equation}

MI(N)LP formulation increases the complexity of the optimisation problem, making the computational time potentially too long for decision-making on the operational horizon, being important in the DOE concept. Even though the HC calculation is more of a planning problem and the larger computational time does not present a problem, it remains the question if the MINLP formulation is able to find a solution in a realistic time. Therefore, we introduce further approaches for determining a DER connection phase without extending the OPF formulation by binary variables.

In real-world LV distribution networks, end-users are unaware of the DER connection phase, i.e., the connection phase is randomly selected for each node to which a DER is connected and each optimisation time step, as shown in \eqref{eq:rand_phase}.
Same as in the formulation with binary variables, we present the assumption that the connection phase potentially changes in different time periods. However, the formulation can be modified by setting the connection phase to be the same in every time period for each node.

\begin{equation}
    q = random\{a,b,c\}, \ \forall n \in N, \ \forall t \in T
    \label{eq:rand_phase}
\end{equation}

Finally, we introduce an additional approach in which binary variables are omitted from the OPF model. In this approach, both maximally and minimally loaded phases of each end-user are determined. In the case of generation, DER is connected to the most loaded phase, while in the case of additional demand, the new load is connected to the least loaded phase. That way, the nodal unbalance will be more balanced than in other cases. Depending on whether the phase-switching is assumed as possible or not, we determine the maximum or minimum loaded phase in each period or the phase with the largest or lowest total electricity demand overall the observed period - 24 hours in this study, as seen in \eqref{eq:max_node_demand}. After the described phase demand calculation, the DER connection phase $g$ is determined with Algorithm \ref{alg:pv_connect}.

\begin{equation}
\begin{gathered}
    D = max/min \{P^{load}_{n,p=a,t}, P^{load}_{n,p=b,t}, P^{load}_{n,p=c,t}\}\\    
    D = max/min \Biggl\{\sum_{t \in T}P^{load}_{n,p=a,t}, \sum_{t \in T}P^{load}_{n,p=b,t}, \sum_{t \in T}P^{load}_{n,p=c,t}\Biggr\}
    \label{eq:max_node_demand}
\end{gathered}
\end{equation}

\begin{algorithm}
\caption{Determining a PV system's variable connection phase}
\begin{algorithmic}
\If {$D = P^{load}_{n,p=a,t} \ or \  D = \sum_{t \in T}P^{load}_{n,p=a,t}$}
    \State {$q = a$}
\ElsIf{$D = P^{load}_{n,p=b,t} \ or \ D = \sum_{t \in T}P^{load}_{n,p=b,t}$}
    \State {$q = b$}
\ElsIf{{$D = P^{load}_{n,p=c,t} \ or \ D = \sum_{t \in T}P^{load}_{n,p=c,t}$}}
    \State {$q = c$}
\EndIf
\end{algorithmic}
\label{alg:pv_connect}
\end{algorithm}

Ensuring that DERs are single-phase connected is the same in both approaches, not relying on extending the OPF formulation by binary variables. In the current-voltage formulation, the real and imaginary part of the current of the non-selected phase is set to be zero \eqref{eq:i_der_zero_rand}, while in the linearised formulation, the DER active power in non-selected phases is set to be zero \eqref{eq:p_der_zero_rand}. As mentioned before, the DER reactive power variable is already set to zero for all phases, so there is no need for additional constraints.

\begin{equation}
    \left(I^{DER}_{n,p \neq q,t}\right)^{re/im} = 0
    \label{eq:i_der_zero_rand}
\end{equation}

\begin{equation}
    P^{DER}_{n,p \neq q,t} = 0
    \label{eq:p_der_zero_rand}
\end{equation}

\subsection{Objective function}
The objective of the given optimisation problem is to calculate the theoretical maximum of installed DER active power, both in cases of static and dynamic export and import limits, as defined with \eqref{eq:obj}. The set of constraints for which the objective function is solved differs based on the problem formulation and scenario, however, the objective function remains unchanged.

\begin{equation}
    obj = \sum_{n \in N} \sum_{p \in P} \sum_{t \in T} P_{n,p,t}^{DER}
    \label{eq:obj}
\end{equation}

Depending on the case and also the OPF formulation, the objective function needs to be slightly modified. The defined objective function is valid in the case of DOE maximisation calculated using the formulation in which binary variables are not introduced and the current-voltage formulation developed as MINLP. Adding variables into the \textit{LinDist3Flow} formulation requires multiplying $P_{n,p,t}^{DER}$ by the binary variable $x_{n,p,t}^{DER}$.

When static limits, i.e., HC, are calculated, there is no need to sum the objective function over time $t$ since the optimisation problem is, in that case, solved for a single time step.

\section{Case Studies} \label{sec:cs}
The developed model is tested in investigating the impact of the connection phase selection on the accuracy and computational complexity of the HC and DOE calculation. To avoid the conclusions that are valid for only one specific case, we test the OPF formulations extended with different techniques to determine the connection phase on two different networks, the residential subnetwork of the synthetic CIGRE LV network \cite{CIGRE_LV} and the real-world residential LV feeder. LV networks are characterized by different topologies and element parameters, but most importantly, by different numbers of nodes and elements, i.e., network sizes. Such a setup allows the investigation of whether the complexity of the network's structure plays an important role in the given problem.

\begin{figure}[htbp]
    \centering
    \includegraphics[width=\textwidth]{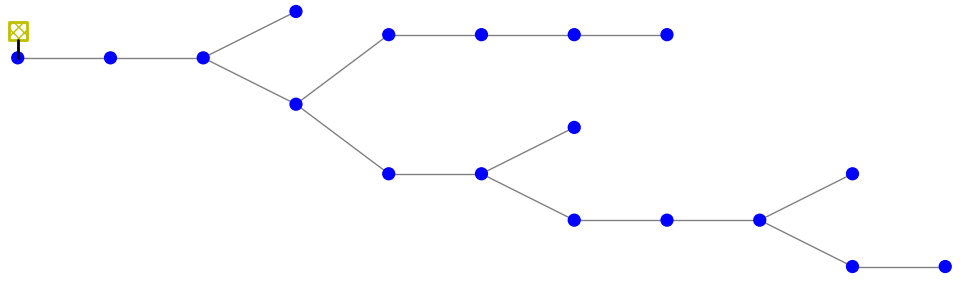}
    \caption{Residential CIGRE LV feeder}
    \label{fig:cigre_lv}
\end{figure}

The residential CIGRE LV feeder consists of 18 0.4 kV nodes and 17 lines. End-users are connected to five nodes, and since they are initially defined as single-phase, their active and reactive power are randomly distributed along the phases. Active and reactive power of loads are defined for only one time period, and this setup is used in the HC calculation. DOE calculation requires demand estimates over time, so the scaled demand curve created from the real-world measurements is used to calculate active and reactive power in different periods. The topology of the feeder is shown in Figure \ref{fig:cigre_lv}.

\begin{figure}[htbp]
    \centering
    \includegraphics[width=\textwidth]{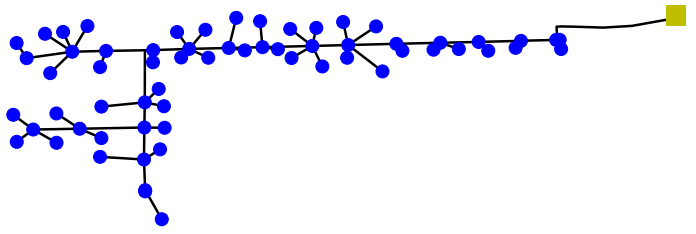}
    \caption{Residential real-world LV feeder}
    \label{fig:rw_lv}
\end{figure}

The real-world residential LV feeder presented in Figure \ref{fig:rw_lv} is part of the Croatian distribution LV network. It contains 64 LV nodes and 63 lines. End-users are connected to 43 nodes and each of them is three-phase connected. Values of active and reactive power in each time period are defined based on the measurements collected from smart meters installed in the network. 

As already mentioned, presented mathematical OPF formulations and their implementation in the ppOPF tool are tested on calculating HC and DOE. In the first case study (CS), HC is calculated for a set of active and reactive power values defined in only one time period. Therefore, differences when using diverse approaches to determine the DER connection phase may not be as significant. Furthermore, the HC calculation is defined as conservative and unnecessarily limits the DER integration. Due to the simultaneity factor being different from one, the initial electricity demand scenario is highly unrealistic. Therefore, in the second CS, we assess dynamic export and import limits, investigating the changes in the accuracy and scalability when a higher number of time periods is set. Case studies are defined as follows:

\begin{itemize}
    \item CS 1-1 HC is calculated in the case of export limits
    \item CS 1-2 HC is calculated in the case of import limits 
    \item CS 2-1 is calculated in the case of dynamic export limits
    \item CS 2-2 is calculated in the case of dynamic import limits
\end{itemize}

Each approach to determining the DER connection phase is defined as a separate scenario. That way, it is possible to investigate the contribution of using binary variables to the increase in total installed DERs power but also to analyse the ratio between the loss of accuracy and the reduction in the computational time when different approximations are introduced. In S1, we use the exact MINLP formulation to calculate theoretically maximum values of both static and dynamic HC. Since the proposed approach has not yet been investigated, scenarios S2-S5 are defined so the presented novel formulation can be quantified. S2 and S3 are the most realistic scenarios since, in most LV networks, the connection phase is randomly selected and unknown to an end user and, in some cases, DSOs. Finally, S4 and S5 are defined to theoretically decrease the nodal unbalance and to see if the assumption maximises the DER installed power and allows avoiding the use of binary variables. Scenarios (S) are defined as follows:

\begin{itemize}
    \item S1: Binary variables are introduced to determine the optimal connection phase of DER
    \item S2: Connection phase of DERs is randomly assigned in every time period
    \item S3: Connection phase of DERs  is randomly assigned but is the same in every time period 
    \item S4: Connection phase changes in each time period and it is equal to the most/least loaded phase
    \item S5: Connection phase is the same in all time periods and it is equal to the one with the highest/lowest summed value of active power
\end{itemize}

For the HC calculation, the pair of scenarios S2-S3 and S4-S5 represents the same scenario since the optimisation problem is solved for only one time step and there is no time-dependent change of the connection phase as in the case of the DOE assessment. 

\section{Results} \label{sec:results}
Analyses in all case studies are done using Python 3.9.16 programming language and the pyomo optimisation framework \cite{Hart2011}. Gurobi, ipopt, and knitro solvers are used to solve defined optimisation problems. Since one of the objectives of the study is to investigate the potential loss of accuracy under certain assumptions, we analyse if different solvers lead to the same solution, i.e., if the objective function value remains the same when the solver is changed. Furthermore, not all solvers are applicable to solving every defined problem. Even though the fastest to solve MILP problems, gurobi cannot successfully solve nonconvex problems. Ipopt is a publicly available solver for solving LP and NLP optimisation problems, however, the solver transforms binary variables into continuous ones, leading to a wrong objective function value. Finally, knitro successfully solves all optimisation problems and different formulations, even though it takes longer time to resolve the MILP problems than gurobi. PC specifications are 11th Gen Intel(R) Core(TM) i7-1165G7 processor and 16.0 GB of RAM.

\subsection{Hosting Capacity}
Table \ref{tab:hc_cs_1_1_cigre} shows the summarised optimisation results of the static export limits calculation in the modelled CIGRE LV network. As can be seen, the computational time values are similar in all scenarios and less than 0.3 seconds in all scenarios, with the exception of Scenario 1, in which the optimisation is solved by the current-voltage OPF formulation, where the computational time is 5.8 seconds, and when solved by the linearised formulation, when the computational time is 2.2 seconds. However, since the formulation in Scenario 1 is MINLP, or MILP, the computational time is not problematic in a planning horizon. 
Furthermore, using different solvers does not cause significant deviations in computational time since it varies in millisecond values. The only significant difference is in the case of using linearised formulation in Scenario 1 when solving the optimisation problem by knitro takes 2.2 seconds and by gurobi 0.3 seconds. Also, the objective function values are the same in all scenarios for the same OPF formulation, no matter which solver was used.
Differences in results are best seen when comparing the objective function values calculated for the same scenarios but with different OPF formulations and when comparing the results of the same formulation but different scenarios. The results show that the objective function value is the highest in Scenario 1, both for NLP and LP OPF formulations. The increase of the objective function is in the range of 2.6\% and 3.9\% when compared to results in scenarios S2-S3 and S4-S5. Also, the values calculated by the nonlinear nonconvex OPF formulations are higher than those calculated by the approximated linearised formulation, meaning that, in this case, defining a more detailed model, considering more physical laws leads to a higher and more accurate solution despite the increase in the computational time. The increase is in milliseconds in scenarios without binary variables, and in Scenario 1, the increase is between 23.6 and 5.5 seconds, depending on the solver used in the LP formulation. However, due to a relatively small increase in computational time, the results show the benefits of using the exact three-phase OPF formulation and suggest its use in planning problems of relatively small LV networks. 
Another important conclusion drawn from the results is that it is not possible to determine the pattern of selecting the optimal connection phase without using binary variables. Randomly selected connection phases lead to a higher objective function value than the selection in which the balance between the phases should be highest in each node. Therefore, the MINLP formulation enables calculating the theoretically maximal aggregated installed DER power in an LV feeder, and for a feeder with a lower number of elements, the formulation can be used in calculating HC, either defined as a planning problem or for the next time step in daily operation.

\begin{table}[htbp]
\centering
\caption{Optimisation results summary - CS 1-1 CIGRE LV network}
\resizebox{\textwidth}{!}{\begin{tabular}{cl|ccc|ccc|}
\cline{3-8}
\multicolumn{1}{l}{}                                  &                                                                                  & \multicolumn{3}{c|}{\textbf{Current-voltage}}                                                & \multicolumn{3}{c|}{\textbf{LinDist3Flow}}                                                   \\ \cline{3-8} 
\multicolumn{1}{l}{}                                  &                                                                                  & \multicolumn{1}{c|}{\textbf{gurobi}} & \multicolumn{1}{c|}{\textbf{ipopt}} & \textbf{knitro} & \multicolumn{1}{c|}{\textbf{gurobi}} & \multicolumn{1}{c|}{\textbf{ipopt}} & \textbf{knitro} \\ \hline
\multicolumn{1}{|c|}{\multirow{2}{*}{\textbf{S1}}}    & \textbf{\begin{tabular}[c]{@{}l@{}}Objective function\\ value (kW)\end{tabular}} & \multicolumn{1}{c|}{N/A}             & \multicolumn{1}{c|}{N/A}            & 334.731         & \multicolumn{1}{c|}{319.940}         & \multicolumn{1}{c|}{N/A}            & 319.940         \\ \cline{2-8} 
\multicolumn{1}{|c|}{}                                & \textbf{\begin{tabular}[c]{@{}l@{}}Computational\\ time (s)\end{tabular}}        & \multicolumn{1}{c|}{N/A}             & \multicolumn{1}{c|}{N/A}            & 5.8092          & \multicolumn{1}{c|}{0.3393}          & \multicolumn{1}{c|}{N/A}            & 2.2051          \\ \hline
\multicolumn{1}{|c|}{\multirow{2}{*}{\textbf{S2-S3}}} & \textbf{\begin{tabular}[c]{@{}l@{}}Objective function\\ value (kW)\end{tabular}} & \multicolumn{1}{c|}{N/A}             & \multicolumn{1}{c|}{322.911}        & 322.911         & \multicolumn{1}{c|}{307.967}         & \multicolumn{1}{c|}{307.967}        & 307.966         \\ \cline{2-8} 
\multicolumn{1}{|c|}{}                                & \textbf{\begin{tabular}[c]{@{}l@{}}Computational\\ time (s)\end{tabular}}        & \multicolumn{1}{c|}{N/A}             & \multicolumn{1}{c|}{0.3002}         & 0.2757          & \multicolumn{1}{c|}{0.2640}          & \multicolumn{1}{c|}{0.1657}         & 0.1858          \\ \hline
\multicolumn{1}{|c|}{\multirow{2}{*}{\textbf{S4-S5}}} & \textbf{\begin{tabular}[c]{@{}l@{}}Objective function\\ value (kW)\end{tabular}} & \multicolumn{1}{c|}{N/A}             & \multicolumn{1}{c|}{325.972}        & 325.972         & \multicolumn{1}{c|}{311.878}         & \multicolumn{1}{c|}{311.878}        & 311.878         \\ \cline{2-8} 
\multicolumn{1}{|c|}{}                                & \textbf{\begin{tabular}[c]{@{}l@{}}Computational\\ time (s)\end{tabular}}        & \multicolumn{1}{c|}{N/A}             & \multicolumn{1}{c|}{0.2871}         & 0.2966          & \multicolumn{1}{c|}{0.2570}          & \multicolumn{1}{c|}{0.1797}         & 0.2091          \\ \hline
\end{tabular}}
\label{tab:hc_cs_1_1_cigre}
\end{table}

The summarised results of the HC calculation in the case of static import limits in the modelled CIGRE LV network are presented in Table \ref{tab:hc_cs_1_2_cigre}. All the conclusions drawn from the analysis of the results when calculating the static export limits remain the same in this case. The computational time values are slightly different, but the general trend remains the same. The values are close to each other, and they differ on a millisecond scale, except in Scenario 1, in which the MINLP formulation takes longer time than MILP, and in the MILP formulation, gurobi solves the optimisation problem almost 2 seconds faster than knitro.
The largest disparity of CS 1-2 results in comparison to CS 1-1 is in each scenario's lower objective function values obtained by the current-voltage and linearised formulations. In this case, the values of the objective function increase when the linearised approximation is used. However, this does not show the benefits or dominance of the linearised formulation over the exact one. On the contrary, it shows that neglecting certain constraints defining important physical laws leads to the loss of accuracy and potentially endangers the safe operation of distribution networks, especially when the network is operated close to its technical limitations.

\begin{table}[htbp]
\centering
\caption{Optimisation results summary - CS 1-2 CIGRE LV network}
\resizebox{\textwidth}{!}{\begin{tabular}{cl|ccc|ccc|}
\cline{3-8}
\multicolumn{1}{l}{}                                  &                                                                                  & \multicolumn{3}{c|}{\textbf{Current-voltage}}                                                & \multicolumn{3}{c|}{\textbf{LinDist3Flow}}                                                   \\ \cline{3-8} 
\multicolumn{1}{l}{}                                  &                                                                                  & \multicolumn{1}{c|}{\textbf{gurobi}} & \multicolumn{1}{c|}{\textbf{ipopt}} & \textbf{knitro} & \multicolumn{1}{c|}{\textbf{gurobi}} & \multicolumn{1}{c|}{\textbf{ipopt}} & \textbf{knitro} \\ \hline
\multicolumn{1}{|c|}{\multirow{2}{*}{\textbf{S1}}}    & \textbf{\begin{tabular}[c]{@{}l@{}}Objective function\\ value (kW)\end{tabular}} & \multicolumn{1}{c|}{N/A}             & \multicolumn{1}{c|}{N/A}            & 187.587         & \multicolumn{1}{c|}{199.188}         & \multicolumn{1}{c|}{N/A}            & 199.188         \\ \cline{2-8} 
\multicolumn{1}{|c|}{}                                & \textbf{\begin{tabular}[c]{@{}l@{}}Computational\\ time (s)\end{tabular}}        & \multicolumn{1}{c|}{N/A}             & \multicolumn{1}{c|}{N/A}            & 6.5747          & \multicolumn{1}{c|}{0.2337}          & \multicolumn{1}{c|}{N/A}            & 2.0887          \\ \hline
\multicolumn{1}{|c|}{\multirow{2}{*}{\textbf{S2-S3}}} & \textbf{\begin{tabular}[c]{@{}l@{}}Objective function\\ value (kW)\end{tabular}} & \multicolumn{1}{c|}{N/A}             & \multicolumn{1}{c|}{173.954}        & 173.954         & \multicolumn{1}{c|}{185.509}         & \multicolumn{1}{c|}{185.509}        & 185.509         \\ \cline{2-8} 
\multicolumn{1}{|c|}{}                                & \textbf{\begin{tabular}[c]{@{}l@{}}Computational\\ time (s)\end{tabular}}        & \multicolumn{1}{c|}{N/A}             & \multicolumn{1}{c|}{0.3085}         & 0.3335          & \multicolumn{1}{c|}{0.3410}          & \multicolumn{1}{c|}{0.1722}         & 0.2641          \\ \hline
\multicolumn{1}{|c|}{\multirow{2}{*}{\textbf{S4-S5}}} & \textbf{\begin{tabular}[c]{@{}l@{}}Objective function\\ value (kW)\end{tabular}} & \multicolumn{1}{c|}{N/A}             & \multicolumn{1}{c|}{160.373}        & 160.373         & \multicolumn{1}{c|}{169.598}         & \multicolumn{1}{c|}{169.598}        & 169.597         \\ \cline{2-8} 
\multicolumn{1}{|c|}{}                                & \textbf{\begin{tabular}[c]{@{}l@{}}Computational\\ time (s)\end{tabular}}        & \multicolumn{1}{c|}{N/A}             & \multicolumn{1}{c|}{0.3689}         & 0.4691          & \multicolumn{1}{c|}{0.3045}          & \multicolumn{1}{c|}{0.1832}         & 0.1887          \\ \hline
\end{tabular}}
\label{tab:hc_cs_1_2_cigre}
\end{table}

Since the CIGRE LV feeder is a network with a relatively small number of elements, the results show that calculating HC does not face scalability issues, even in the case of the computationally complex MINLP formulation. Therefore, the same calculations are performed on a model of a larger real-world residential LV feeder.
The increase in the computational time when using NLP OPF formulation is much more emphasised in this case, even though the maximum computational time of 1.75 seconds in S2-S3 does not present an issue in the planning and operation of distribution networks. 
The conclusions related to the objective function value remain the same as in the CIGRE LV network; the current-voltage formulation ensures higher values in comparison to the linearised formulation. The difference between the values is also similar to the one in the CIGRE LV network, meaning that it is not possible to directly conclude the correlation between the network size and the approximation accuracy.
The biggest impact of the networks' size is on the computational time in S1, a scenario in which binary variables are introduced in the OPF formulation in order to determine the optimal connection phase. The only case in which optimisation is performed within acceptable time is the simulation in which the gurobi solver is used in the \textit{LinDist3Flow} formulation when the calculation is performed in over 2 seconds. However, the optimality gap is set to 0.15\% since the initial attempts without defining the optimality gap did not result in a successful calculation. An even higher optimality gap of 0.5\% needed to be defined when the knitro solver was used in both nonlinear and linear formulations. The optimal solution was found in a time of over 5 minutes with the linear formulation and almost 8 minutes with the exact current-voltage formulation. In the case of the MILP formulation, it is possible to use the gurobi solver and successfully resolve the computational complexity. However, in the MINLP formulation, the computational time presents a potential problem in calculating DER HC. The increase of the computational accuracy and the objective function value, i.e., the installed DER power, show the benefits of the MINLP formulation. Since in CS1, it is applied in the planning phase, the computational time of 8 minutes does not present a problem for a DSO and the results show that the exact formulation can be used in real-world networks.

\begin{table}[htbp]
\centering
\caption{Optimisation results summary - CS 1-1 Real-world LV feeder}
\resizebox{\textwidth}{!}{\begin{tabular}{cl|ccc|ccc|}
\cline{3-8}
\multicolumn{1}{l}{}                                  &                                                                                  & \multicolumn{3}{c|}{\textbf{Current-voltage}}                                                & \multicolumn{3}{c|}{\textbf{LinDist3Flow}}                                                   \\ \cline{3-8} 
\multicolumn{1}{l}{}                                  &                                                                                  & \multicolumn{1}{c|}{\textbf{gurobi}} & \multicolumn{1}{c|}{\textbf{ipopt}} & \textbf{knitro} & \multicolumn{1}{c|}{\textbf{gurobi}} & \multicolumn{1}{c|}{\textbf{ipopt}} & \textbf{knitro} \\ \hline
\multicolumn{1}{|c|}{\multirow{2}{*}{\textbf{S1}}}    & \textbf{\begin{tabular}[c]{@{}l@{}}Objective function\\ value (kW)\end{tabular}} & \multicolumn{1}{c|}{N/A}             & \multicolumn{1}{c|}{N/A}            & 474.537         & \multicolumn{1}{c|}{454.959}         & \multicolumn{1}{c|}{N/A}            & 453.961         \\ \cline{2-8} 
\multicolumn{1}{|c|}{}                                & \textbf{\begin{tabular}[c]{@{}l@{}}Computational\\ time (s)\end{tabular}}        & \multicolumn{1}{c|}{N/A}             & \multicolumn{1}{c|}{N/A}            & 478.8724        & \multicolumn{1}{c|}{2.2174}          & \multicolumn{1}{c|}{N/A}            & 304.7852        \\ \hline
\multicolumn{1}{|c|}{\multirow{2}{*}{\textbf{S2-S3}}} & \textbf{\begin{tabular}[c]{@{}l@{}}Objective function\\ value (kW)\end{tabular}} & \multicolumn{1}{c|}{N/A}             & \multicolumn{1}{c|}{258.530}        & 258.530         & \multicolumn{1}{c|}{256.953}         & \multicolumn{1}{c|}{256.953}        & 256.944         \\ \cline{2-8} 
\multicolumn{1}{|c|}{}                                & \textbf{\begin{tabular}[c]{@{}l@{}}Computational\\ time (s)\end{tabular}}        & \multicolumn{1}{c|}{N/A}             & \multicolumn{1}{c|}{1.7511}         & 1.0185          & \multicolumn{1}{c|}{0.3112}          & \multicolumn{1}{c|}{0.6151}         & 0.3526          \\ \hline
\multicolumn{1}{|c|}{\multirow{2}{*}{\textbf{S4-S5}}} & \textbf{\begin{tabular}[c]{@{}l@{}}Objective function\\ value (kW)\end{tabular}} & \multicolumn{1}{c|}{N/A}             & \multicolumn{1}{c|}{307.428}        & 307.428         & \multicolumn{1}{c|}{293.927}         & \multicolumn{1}{c|}{293.927}        & 293.922         \\ \cline{2-8} 
\multicolumn{1}{|c|}{}                                & \textbf{\begin{tabular}[c]{@{}l@{}}Computational\\ time (s)\end{tabular}}        & \multicolumn{1}{c|}{N/A}             & \multicolumn{1}{c|}{1.6289}         & 1.0678          & \multicolumn{1}{c|}{0.4582}          & \multicolumn{1}{c|}{0.4273}         & 0.3301          \\ \hline
\end{tabular}}
\label{tab:hc_cs_1_1_rw_feeder}
\end{table}

Calculated static import limits in CS 2-2 are always equal to zero, i.e., it is not possible to add additional demand in the modelled network. According to the HC definition, we consider the highest electricity demand value for each end-user in the optimisation problem. The optimisation model used to calculate the network's HC in the case of import is infeasible due to the violation of voltage constraints even without added DERs. The analysis shows that with the defined setup, voltage magnitude goes as low as 0.83 p.u., which is lower than the defined threshold value of 0.9 p.u. These results show that the traditional approach to the HC calculation is conservative and unsustainable since it prevents and slows down the installation of new DERs. 

\subsection{Dynamic Operating Envelopes}
Since the results of the HC calculation show that, in some cases, it is a conservative approach limiting the share of DERs, we present a detailed investigation of the accuracy and scalability of the novel DOE concept.  DOE is calculated from the operational perspective, assessing the dynamic export and import limits for the day ahead. We observe a single-day, 15-minute interval of electricity demand measurements, i.e., there is a total of 96 measurements for each end-user defined as day-ahead demand estimates. With the number of time steps in the optimisation problem significantly larger than in the case of the HC assessment, the differences between different approaches and OPF formulations are expected to be much more significant. Due to the fact that the benefits of DOE, i.e., time-varying HC, are already known and have been well-investigated, the quantitative comparison between the static HC values calculated in the planning phase and dynamic HC values calculated in the operational phase is not the focus of this study. Instead, we investigate how the increase in the number of periods impacts the accuracy and scalability of the defined problem.

Table \ref{tab:hc_cs_2_1_cigre_lv} presents summarised results of calculating dynamic export limits in the CIGRE LV network. In this case, there are no scalability issues, even in S1, in the case of MINLP formulation, which takes more than 10 minutes to solve the defined optimisation problem. However, given the definition of the DOE calculation as a day-ahead concept in this study, this is not a concerning value and guarantees the increase of the objective function value for 4.9\% when compared to the results obtained using the MILP formulation.
The results in other scenarios are in line with those in the HC calculation. It is not possible to determine a direct correlation between the DER connection phase and the objective function value. For example, one could expect the best results in S4, in which the connection phase changes, so the balance between the phases decreases. Yet, the objective function value is higher in S5, in which the connection phase is static and does not change over time. This highlights the importance of finding the optimal selection phase since it can increase the daily export limits up to 7 MW on an aggregated level.
In the results calculated by the nonlinear nonconvex three-phase OPF formulation, it is still not possible to determine whether the performance of the ipopt solver is faster than the one of knitro or vice versa since, in some scenarios, ipopt is faster while in others, knitro is. However, the performances of solvers in the case of a linearised formulation without binary variables are easier to compare due to the clear pattern in which the ipopt solver is the fastest while gurobi is the slowest. Even though the differences are up to more than 6 seconds, they are not the decisive factor in selecting the solver since the problem is calculated for the dynamic export limits the next day.

\begin{table}[htbp]
\centering
\caption{Optimisation results summary - CS 2-1 CIGRE LV network}
\resizebox{\textwidth}{!}{\begin{tabular}{cl|ccc|ccc|}
\cline{3-8}
\multicolumn{1}{l}{}                               &                                                                                  & \multicolumn{3}{c|}{\textbf{Current-voltage}}                                                & \multicolumn{3}{c|}{\textbf{LinDist3Flow}}                                                   \\ \cline{3-8} 
\multicolumn{1}{l}{}                               &                                                                                  & \multicolumn{1}{c|}{\textbf{gurobi}} & \multicolumn{1}{c|}{\textbf{ipopt}} & \textbf{knitro} & \multicolumn{1}{c|}{\textbf{gurobi}} & \multicolumn{1}{c|}{\textbf{ipopt}} & \textbf{knitro} \\ \hline
\multicolumn{1}{|c|}{\multirow{2}{*}{\textbf{S1}}} & \textbf{\begin{tabular}[c]{@{}l@{}}Objective function\\ value (kW)\end{tabular}} & \multicolumn{1}{c|}{N/A}             & \multicolumn{1}{c|}{N/A}            & 27677.190       & \multicolumn{1}{c|}{26387.928}       & \multicolumn{1}{c|}{N/A}            & 26387.928       \\ \cline{2-8} 
\multicolumn{1}{|c|}{}                             & \textbf{\begin{tabular}[c]{@{}l@{}}Computational\\ time (s)\end{tabular}}        & \multicolumn{1}{c|}{N/A}             & \multicolumn{1}{c|}{N/A}            & 620.5988        & \multicolumn{1}{c|}{24.2337}         & \multicolumn{1}{c|}{N/A}            & 314.6566        \\ \hline
\multicolumn{1}{|c|}{\multirow{2}{*}{\textbf{S2}}} & \textbf{\begin{tabular}[c]{@{}l@{}}Objective function\\ value (kW)\end{tabular}} & \multicolumn{1}{c|}{N/A}             & \multicolumn{1}{c|}{22360.252}      & 22360.249       & \multicolumn{1}{c|}{21548.757}       & \multicolumn{1}{c|}{21548.757}      & 21548.744       \\ \cline{2-8} 
\multicolumn{1}{|c|}{}                             & \textbf{\begin{tabular}[c]{@{}l@{}}Computational\\ time (s)\end{tabular}}        & \multicolumn{1}{c|}{N/A}             & \multicolumn{1}{c|}{27.6885}        & 29.2142         & \multicolumn{1}{c|}{18.1022}         & \multicolumn{1}{c|}{12.4514}        & 15.7861         \\ \hline
\multicolumn{1}{|c|}{\multirow{2}{*}{\textbf{S3}}} & \textbf{\begin{tabular}[c]{@{}l@{}}Objective function\\ value (kW)\end{tabular}} & \multicolumn{1}{c|}{N/A}             & \multicolumn{1}{c|}{20593.537}      & 20593.537       & \multicolumn{1}{c|}{19798.485}       & \multicolumn{1}{c|}{19798.485}      & 19798.481       \\ \cline{2-8} 
\multicolumn{1}{|c|}{}                             & \textbf{\begin{tabular}[c]{@{}l@{}}Computational\\ time (s)\end{tabular}}        & \multicolumn{1}{c|}{N/A}             & \multicolumn{1}{c|}{28.5668}        & 28.6674         & \multicolumn{1}{c|}{17.4586}         & \multicolumn{1}{c|}{12.0894}        & 14.7833         \\ \hline
\multicolumn{1}{|c|}{\multirow{2}{*}{\textbf{S4}}} & \textbf{\begin{tabular}[c]{@{}l@{}}Objective function\\ value (kW)\end{tabular}} & \multicolumn{1}{c|}{N/A}             & \multicolumn{1}{c|}{23804.624}      & 23804.622       & \multicolumn{1}{c|}{22803.493}       & \multicolumn{1}{c|}{22803.493}      & 22803.482       \\ \cline{2-8} 
\multicolumn{1}{|c|}{}                             & \textbf{\begin{tabular}[c]{@{}l@{}}Computational\\ time (s)\end{tabular}}        & \multicolumn{1}{c|}{N/A}             & \multicolumn{1}{c|}{28.3143}        & 27.4034         & \multicolumn{1}{c|}{17.3873}         & \multicolumn{1}{c|}{11.4203}        & 14.5343         \\ \hline
\multicolumn{1}{|c|}{\multirow{2}{*}{\textbf{S5}}} & \textbf{\begin{tabular}[c]{@{}l@{}}Objective function\\ value (kW)\end{tabular}} & \multicolumn{1}{c|}{N/A}             & \multicolumn{1}{c|}{26639.924}      & 26639.923       & \multicolumn{1}{c|}{25386.216}       & \multicolumn{1}{c|}{25386.217}      & 25386.216       \\ \cline{2-8} 
\multicolumn{1}{|c|}{}                             & \textbf{\begin{tabular}[c]{@{}l@{}}Computational\\ time (s)\end{tabular}}        & \multicolumn{1}{c|}{N/A}             & \multicolumn{1}{c|}{24.8822}        & 29.0250         & \multicolumn{1}{c|}{18.3244}         & \multicolumn{1}{c|}{12.2882}        & 14.5583         \\ \hline
\end{tabular}}
\label{tab:hc_cs_2_1_cigre_lv}
\end{table}

In the case of calculating dynamic import limits, there are no significant qualitative differences when compared to results in CS 2-1. The computational time needed to find the optimal solution is similar in all scenarios, and calculated import limits are lower than export ones. Such a conclusion is in line with higher initial electricity demand in a network, meaning that the available range is lower and the threshold values are reached faster. 
In all scenarios, the objective function values calculated by \textit{LinDist3Flow} are higher than the values calculated by the current-voltage nonlinear formulation. Such results do not suggest that the linearised formulation should be used in the specific optimisation problem but that the introduced approximation leads to higher voltage values, allowing higher additional demand before the lower voltage bound is reached.
Again, the objective function values calculated when using binary variables are the highest, even though the difference between the results in S1 and second best scenario S5 are around 130 - 145 kW, depending on the OPF formulation. Computational time in the MINLP is almost 20 times as high as in other scenarios, both when using LP and NLP formulations. However, high computational time is not limiting in the day-ahead DOE calculation, and the approach proposed in S1 can be used by DSO to maximise import limits in every time period and also to ensure the safe operation of a network.

\begin{table}[htbp]
\centering
\caption{Optimisation results summary - CS 2-2 CIGRE LV network}
\resizebox{\textwidth}{!}{\begin{tabular}{cl|ccc|ccc|}
\cline{3-8}
\multicolumn{1}{l}{}                               &                                                                                  & \multicolumn{3}{c|}{\textbf{Current-voltage}}                                                & \multicolumn{3}{c|}{\textbf{LinDist3Flow}}                                                   \\ \cline{3-8} 
\multicolumn{1}{l}{}                               &                                                                                  & \multicolumn{1}{c|}{\textbf{gurobi}} & \multicolumn{1}{c|}{\textbf{ipopt}} & \textbf{knitro} & \multicolumn{1}{c|}{\textbf{gurobi}} & \multicolumn{1}{c|}{\textbf{ipopt}} & \textbf{knitro} \\ \hline
\multicolumn{1}{|c|}{\multirow{2}{*}{\textbf{S1}}} & \textbf{\begin{tabular}[c]{@{}l@{}}Objective function\\ value (kW)\end{tabular}} & \multicolumn{1}{c|}{N/A}             & \multicolumn{1}{c|}{N/A}            & 20704.384       & \multicolumn{1}{c|}{21918.746}       & \multicolumn{1}{c|}{N/A}            & 21918.467       \\ \cline{2-8} 
\multicolumn{1}{|c|}{}                             & \textbf{\begin{tabular}[c]{@{}l@{}}Computational\\ time (s)\end{tabular}}        & \multicolumn{1}{c|}{N/A}             & \multicolumn{1}{c|}{N/A}            & 540.2693        & \multicolumn{1}{c|}{27.2956}         & \multicolumn{1}{c|}{N/A}            & 295.4141        \\ \hline
\multicolumn{1}{|c|}{\multirow{2}{*}{\textbf{S2}}} & \textbf{\begin{tabular}[c]{@{}l@{}}Objective function\\ value (kW)\end{tabular}} & \multicolumn{1}{c|}{N/A}             & \multicolumn{1}{c|}{16827.670}      & 16827.670       & \multicolumn{1}{c|}{17907.450}       & \multicolumn{1}{c|}{17907.450}      & 17907.480       \\ \cline{2-8} 
\multicolumn{1}{|c|}{}                             & \textbf{\begin{tabular}[c]{@{}l@{}}Computational\\ time (s)\end{tabular}}        & \multicolumn{1}{c|}{N/A}             & \multicolumn{1}{c|}{25.3544}        & 26.2924         & \multicolumn{1}{c|}{18.1898}         & \multicolumn{1}{c|}{12.2420}        & 15.8128         \\ \hline
\multicolumn{1}{|c|}{\multirow{2}{*}{\textbf{S3}}} & \textbf{\begin{tabular}[c]{@{}l@{}}Objective function\\ value (kW)\end{tabular}} & \multicolumn{1}{c|}{N/A}             & \multicolumn{1}{c|}{15886.459}      & 15886.459       & \multicolumn{1}{c|}{16419.804}       & \multicolumn{1}{c|}{16419.804}      & 16419.784       \\ \cline{2-8} 
\multicolumn{1}{|c|}{}                             & \textbf{\begin{tabular}[c]{@{}l@{}}Computational\\ time (s)\end{tabular}}        & \multicolumn{1}{c|}{N/A}             & \multicolumn{1}{c|}{30.8036}        & 30.1312         & \multicolumn{1}{c|}{18.6670}         & \multicolumn{1}{c|}{12.3382}        & 14.6782         \\ \hline
\multicolumn{1}{|c|}{\multirow{2}{*}{\textbf{S4}}} & \textbf{\begin{tabular}[c]{@{}l@{}}Objective function\\ value (kW)\end{tabular}} & \multicolumn{1}{c|}{N/A}             & \multicolumn{1}{c|}{16308.798}      & 16305.794       & \multicolumn{1}{c|}{17603.561}       & \multicolumn{1}{c|}{17603.561}      & 17603.551       \\ \cline{2-8} 
\multicolumn{1}{|c|}{}                             & \textbf{\begin{tabular}[c]{@{}l@{}}Computational\\ time (s)\end{tabular}}        & \multicolumn{1}{c|}{N/A}             & \multicolumn{1}{c|}{24.3836}        & 27.8321         & \multicolumn{1}{c|}{17.8537}         & \multicolumn{1}{c|}{12.5318}        & 15.7433         \\ \hline
\multicolumn{1}{|c|}{\multirow{2}{*}{\textbf{S5}}} & \textbf{\begin{tabular}[c]{@{}l@{}}Objective function\\ value (kW)\end{tabular}} & \multicolumn{1}{c|}{N/A}             & \multicolumn{1}{c|}{20573.919}      & 20573.918       & \multicolumn{1}{c|}{21774.764}       & \multicolumn{1}{c|}{21774.764}      & 21774.753       \\ \cline{2-8} 
\multicolumn{1}{|c|}{}                             & \textbf{\begin{tabular}[c]{@{}l@{}}Computational\\ time (s)\end{tabular}}        & \multicolumn{1}{c|}{N/A}             & \multicolumn{1}{c|}{24.6185}        & 28.9288         & \multicolumn{1}{c|}{19.1228}         & \multicolumn{1}{c|}{12.6800}        & 16.0297         \\ \hline
\end{tabular}}
\label{tab:hc_cs_2_2_cigre_lv}
\end{table}

The complexity of the optimisation problem increases with the size of a modelled network and the number of time intervals, as seen from the results in Table \ref{tab:hc_cs_2_1_rw_feeder} and Table \ref{tab:hc_cs_2_2_rw_feeder}, presenting the summarised optimisation results in the case of calculating dynamic export and import limits in a real-world LV feeder.
The main conclusions of the analysis of the CIGRE LV network are also valid in this case. Without uplifting the OPF formulation to the MI(N)LP, it is hard to determine a direct correlation between selecting the connection phase and the change in the objective function value. There is no clear pattern showing that the increase in total demand balance between phases enables higher electricity export or import, nor that the stochastic connection phase change means higher objective function value. The results in scenarios other than S1 are different than in the CIGRE LV network, i.e., the second-best scenario is not the same as in the CIGRE network, despite all scenarios being created in the same way. This only contributes to the conclusion that it is not possible to improve optimisation results without ensuring optimality by introducing binary variables.
The complexity of the calculation by extending the DOE calculation to more than one time period is also seen in the optimality gap set in the MI(N)LP formulations. Without setting the optimality gap, i.e., as in the DOE calculation for the CIGRE network, optimisation takes too long time, and it remains the question of when and if the process will be finished. In the HC assessment in the real-world LV feeder, the optimality gap in the case of export was set to 0.15\% when solving the problem by gurobi, and the same gap remains in this case. By setting this gap, computational time is around 7 minutes, which is acceptable in the DOE concept defined in this study. In CS 2-2, setting the optimality gap in the MILP formulation was not needed, and the optimisation process duration was less than 40 seconds, meaning that the linearised problem is not as complex and the gurobi solver is able to solve the problem relatively fast. The situation changes when using the knitro solver since it does not successfully solve the complex problem defined in the real-world LV network optimisation. In CS 2-1, i.e., in the case of export, MINLP and MILP optimisation problems could not be solved after more than 12 hours, even when the optimality gap was set as high as 5\%. Setting the optimality gap so high or even higher would lead to a suboptimal solution, and it remains the question of whether such an approach is more accurate than using the MILP formulation or some of the NLP approximations.
In the MINLP formulation in CS 2-2, the optimisation problem was solved in just over 1 hour with an optimality gap of 3\%. However, the suboptimal solution in this case is lower than the one calculated by the gurobi solver. Therefore, the accuracy is decreased less by introducing the linearised OPF formulation than by increasing the optimality gap.

\begin{table}[htbp]
\centering
\caption{Optimisation results summary - CS 2-1 Real-world LV feeder}
\resizebox{\textwidth}{!}{\begin{tabular}{cl|ccc|ccc|}
\cline{3-8}
\multicolumn{1}{l}{}                               &                                                                                  & \multicolumn{3}{c|}{\textbf{Current-voltage}}                                                & \multicolumn{3}{c|}{\textbf{LinDist3Flow}}                                                   \\ \cline{3-8} 
\multicolumn{1}{l}{}                               &                                                                                  & \multicolumn{1}{c|}{\textbf{gurobi}} & \multicolumn{1}{c|}{\textbf{ipopt}} & \textbf{knitro} & \multicolumn{1}{c|}{\textbf{gurobi}} & \multicolumn{1}{c|}{\textbf{ipopt}} & \textbf{knitro} \\ \hline
\multicolumn{1}{|c|}{\multirow{2}{*}{\textbf{S1}}} & \textbf{\begin{tabular}[c]{@{}l@{}}Objective function\\ value (kW)\end{tabular}} & \multicolumn{1}{c|}{N/A}             & \multicolumn{1}{c|}{N/A}            & \textbf{---}             & \multicolumn{1}{c|}{46705.731}       & \multicolumn{1}{c|}{N/A}            & 44010.532             \\ \cline{2-8} 
\multicolumn{1}{|c|}{}                             & \textbf{\begin{tabular}[c]{@{}l@{}}Computational\\ time (s)\end{tabular}}        & \multicolumn{1}{c|}{N/A}             & \multicolumn{1}{c|}{N/A}            & $>$ 12 hours             & \multicolumn{1}{c|}{425.651}         & \multicolumn{1}{c|}{N/A}            & 9647.0992             \\ \hline
\multicolumn{1}{|c|}{\multirow{2}{*}{\textbf{S2}}} & \textbf{\begin{tabular}[c]{@{}l@{}}Objective function\\ value (kW)\end{tabular}} & \multicolumn{1}{c|}{N/A}             & \multicolumn{1}{c|}{40958.437}      & 40957.248       & \multicolumn{1}{c|}{39351.058}       & \multicolumn{1}{c|}{39351.049}      & 39350.407       \\ \cline{2-8} 
\multicolumn{1}{|c|}{}                             & \textbf{\begin{tabular}[c]{@{}l@{}}Computational\\ time (s)\end{tabular}}        & \multicolumn{1}{c|}{N/A}             & \multicolumn{1}{c|}{221.7191}       & 133.4251        & \multicolumn{1}{c|}{32.2358}         & \multicolumn{1}{c|}{46.5950}        & 35.9461         \\ \hline
\multicolumn{1}{|c|}{\multirow{2}{*}{\textbf{S3}}} & \textbf{\begin{tabular}[c]{@{}l@{}}Objective function\\ value (kW)\end{tabular}} & \multicolumn{1}{c|}{N/A}             & \multicolumn{1}{c|}{48549.955}      & 48547.990       & \multicolumn{1}{c|}{46518.825}       & \multicolumn{1}{c|}{46815.817}      & 46815.172       \\ \cline{2-8} 
\multicolumn{1}{|c|}{}                             & \textbf{\begin{tabular}[c]{@{}l@{}}Computational\\ time (s)\end{tabular}}        & \multicolumn{1}{c|}{N/A}             & \multicolumn{1}{c|}{244.7531}       & 147.0650        & \multicolumn{1}{c|}{31.0893}         & \multicolumn{1}{c|}{43.4804}        & 29.1881         \\ \hline
\multicolumn{1}{|c|}{\multirow{2}{*}{\textbf{S4}}} & \textbf{\begin{tabular}[c]{@{}l@{}}Objective function\\ value (kW)\end{tabular}} & \multicolumn{1}{c|}{N/A}             & \multicolumn{1}{c|}{41677.909}      & 41676.756       & \multicolumn{1}{c|}{40038.749}       & \multicolumn{1}{c|}{40038.741}      & 40038.162       \\ \cline{2-8} 
\multicolumn{1}{|c|}{}                             & \textbf{\begin{tabular}[c]{@{}l@{}}Computational\\ time (s)\end{tabular}}        & \multicolumn{1}{c|}{N/A}             & \multicolumn{1}{c|}{216.0557}       & 121.0496        & \multicolumn{1}{c|}{38.9668}         & \multicolumn{1}{c|}{39.2710}        & 30.3841         \\ \hline
\multicolumn{1}{|c|}{\multirow{2}{*}{\textbf{S5}}} & \textbf{\begin{tabular}[c]{@{}l@{}}Objective function\\ value (kW)\end{tabular}} & \multicolumn{1}{c|}{N/A}             & \multicolumn{1}{c|}{34110.139}      & 34109.999       & \multicolumn{1}{c|}{32797.115}       & \multicolumn{1}{c|}{32797.106}      & 32796.741       \\ \cline{2-8} 
\multicolumn{1}{|c|}{}                             & \textbf{\begin{tabular}[c]{@{}l@{}}Computational\\ time (s)\end{tabular}}        & \multicolumn{1}{c|}{N/A}             & \multicolumn{1}{c|}{195.4838}       & 106.7260        & \multicolumn{1}{c|}{35.0174}         & \multicolumn{1}{c|}{39.2289}        & 30.5000         \\ \hline
\end{tabular}}
\label{tab:hc_cs_2_1_rw_feeder}
\end{table}

\begin{table}[htbp]
\centering
\caption{Optimisation results summary - CS 2-2 Real-world LV feeder}
\resizebox{\textwidth}{!}{\begin{tabular}{cl|ccc|ccc|}
\cline{3-8}
\multicolumn{1}{l}{}                               &                                                                                  & \multicolumn{3}{c|}{\textbf{Current-voltage}}                                                & \multicolumn{3}{c|}{\textbf{LinDist3Flow}}                                                   \\ \cline{3-8} 
\multicolumn{1}{l}{}                               &                                                                                  & \multicolumn{1}{c|}{\textbf{gurobi}} & \multicolumn{1}{c|}{\textbf{ipopt}} & \textbf{knitro} & \multicolumn{1}{c|}{\textbf{gurobi}} & \multicolumn{1}{c|}{\textbf{ipopt}} & \textbf{knitro} \\ \hline
\multicolumn{1}{|c|}{\multirow{2}{*}{\textbf{S1}}} & \textbf{\begin{tabular}[c]{@{}l@{}}Objective function\\ value (kW)\end{tabular}} & \multicolumn{1}{c|}{N/A}             & \multicolumn{1}{c|}{N/A}            & 29653.896             & \multicolumn{1}{c|}{32764.107}       & \multicolumn{1}{c|}{N/A}            & \textbf{---}             \\ \cline{2-8} 
\multicolumn{1}{|c|}{}                             & \textbf{\begin{tabular}[c]{@{}l@{}}Computational\\ time (s)\end{tabular}}        & \multicolumn{1}{c|}{N/A}             & \multicolumn{1}{c|}{N/A}            & 4140.3297             & \multicolumn{1}{c|}{39.5033}         & \multicolumn{1}{c|}{N/A}            & $>$ 12 hours             \\ \hline
\multicolumn{1}{|c|}{\multirow{2}{*}{\textbf{S2}}} & \textbf{\begin{tabular}[c]{@{}l@{}}Objective function\\ value (kW)\end{tabular}} & \multicolumn{1}{c|}{N/A}             & \multicolumn{1}{c|}{25458.654}      & 25458.616       & \multicolumn{1}{c|}{27112.538}       & \multicolumn{1}{c|}{27112.571}      & 27112.364       \\ \cline{2-8} 
\multicolumn{1}{|c|}{}                             & \textbf{\begin{tabular}[c]{@{}l@{}}Computational\\ time (s)\end{tabular}}        & \multicolumn{1}{c|}{N/A}             & \multicolumn{1}{c|}{103.9111}       & 91.7890         & \multicolumn{1}{c|}{31.4614}         & \multicolumn{1}{c|}{38.1413}        & 33.1438         \\ \hline
\multicolumn{1}{|c|}{\multirow{2}{*}{\textbf{S3}}} & \textbf{\begin{tabular}[c]{@{}l@{}}Objective function\\ value (kW)\end{tabular}} & \multicolumn{1}{c|}{N/A}             & \multicolumn{1}{c|}{30555.071}      & 30554.999       & \multicolumn{1}{c|}{32650.5537}      & \multicolumn{1}{c|}{32650.613}      & 32650.413       \\ \cline{2-8} 
\multicolumn{1}{|c|}{}                             & \textbf{\begin{tabular}[c]{@{}l@{}}Computational\\ time (s)\end{tabular}}        & \multicolumn{1}{c|}{N/A}             & \multicolumn{1}{c|}{93.8794}        & 72.8601         & \multicolumn{1}{c|}{31.5980}         & \multicolumn{1}{c|}{39.0850}        & 34.0129         \\ \hline
\multicolumn{1}{|c|}{\multirow{2}{*}{\textbf{S4}}} & \textbf{\begin{tabular}[c]{@{}l@{}}Objective function\\ value (kW)\end{tabular}} & \multicolumn{1}{c|}{N/A}             & \multicolumn{1}{c|}{25847.904}      & 25847.863       & \multicolumn{1}{c|}{27547.3709}      & \multicolumn{1}{c|}{27547.406}      & 27547.110       \\ \cline{2-8} 
\multicolumn{1}{|c|}{}                             & \textbf{\begin{tabular}[c]{@{}l@{}}Computational\\ time (s)\end{tabular}}        & \multicolumn{1}{c|}{N/A}             & \multicolumn{1}{c|}{95.8291}        & 85.5038         & \multicolumn{1}{c|}{34.3709}         & \multicolumn{1}{c|}{35.6906}        & 30.8460         \\ \hline
\multicolumn{1}{|c|}{\multirow{2}{*}{\textbf{S5}}} & \textbf{\begin{tabular}[c]{@{}l@{}}Objective function\\ value (kW)\end{tabular}} & \multicolumn{1}{c|}{N/A}             & \multicolumn{1}{c|}{22940.354}      & 22940.323       & \multicolumn{1}{c|}{24505.455}       & \multicolumn{1}{c|}{24505.479}      & 24505.327       \\ \cline{2-8} 
\multicolumn{1}{|c|}{}                             & \textbf{\begin{tabular}[c]{@{}l@{}}Computational\\ time (s)\end{tabular}}        & \multicolumn{1}{c|}{N/A}             & \multicolumn{1}{c|}{102.4541}       & 74.3547         & \multicolumn{1}{c|}{31.1816}         & \multicolumn{1}{c|}{34.0570}        & 29.3608         \\ \hline
\end{tabular}}
\label{tab:hc_cs_2_2_rw_feeder}
\end{table}

\subsection{Accuracy Analysis}
Since the knitro is the only solver able to solve the defined objective problem in all scenarios and case studies, the accuracy of different formulations is investigated only for the results calculated by knitro. Furthermore, the summarised results show there are no significant differences in objective function values when using gurobi or ipopt. The accuracy of different formulations is tested by comparing per-phase voltage magnitude calculated by NLP and LP formulations in different scenarios. The DER export and import power values are compared in previous analyses of the objective function and computational time values.
Also, voltage magnitude values are compared only for the HC calculation. The results show that general conclusions in the HC calculation are also valid in the DOE calculation. The HC is calculated for a single time period, and the visualization of these results is much more understandable. Since the modelled real-world network contains more nodes and branches, results calculated for this network are used for the accuracy analysis when calculating export limits. However, as mentioned in the previous analysis, the import values could not be calculated in the defined real-world network setup. Therefore, the analysis in the case of import values is performed for the CIGRE LV network.

\begin{figure}
    \centering
    \includegraphics[width=\textwidth]{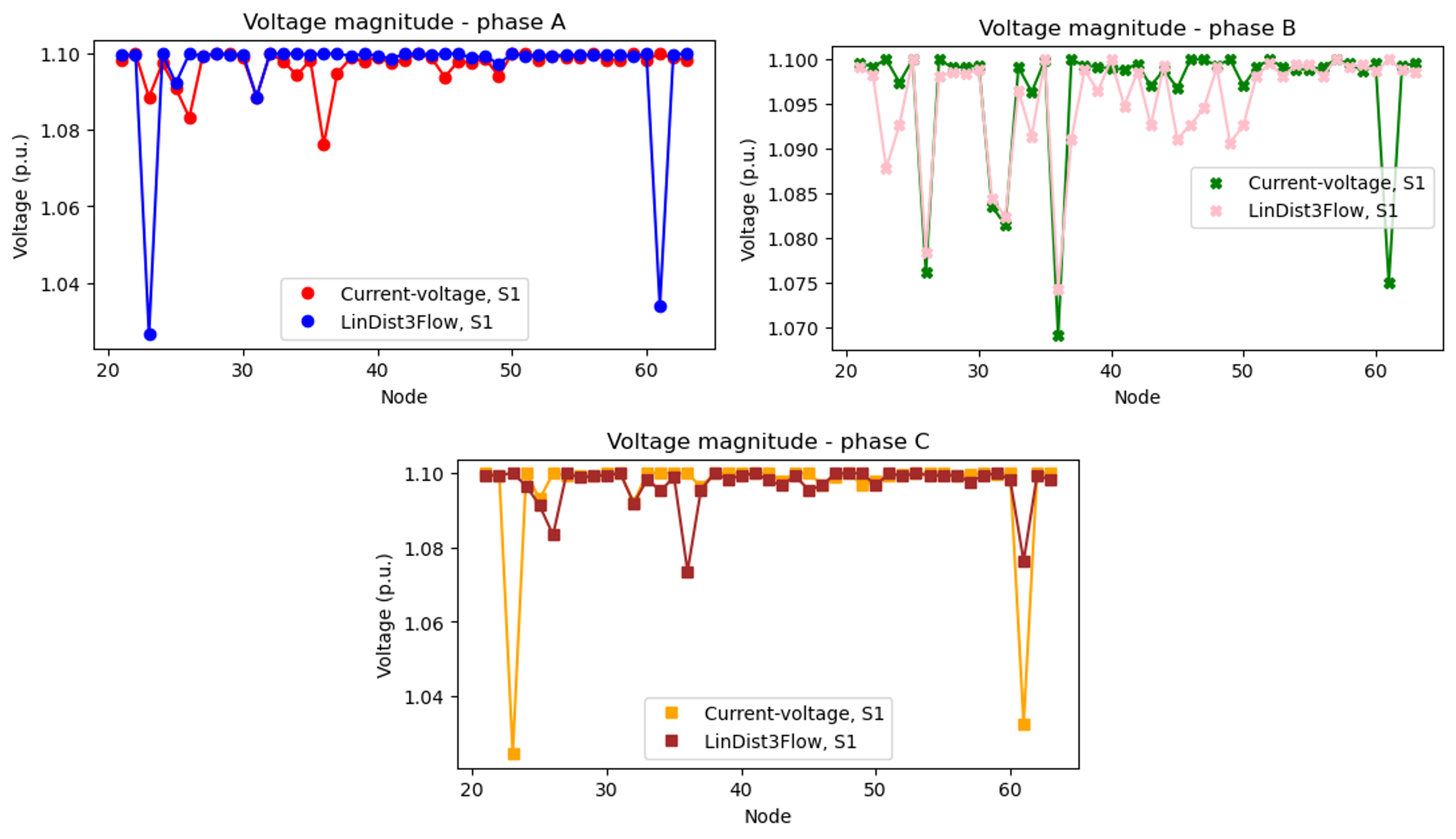}
    \caption{Voltage magnitude in HC calculation, CS 1-1, S1 - Real-world feeder}
    \label{fig:hc_voltage_cs1_s1}
\end{figure}

Figure \ref{fig:hc_voltage_cs1_s1} shows per-phase voltage magnitudes in CS 1-1, in Scenario 1, i.e., it shows the per-phase voltage magnitude in case of calculating static export limits in the real-world LV feeder. For the purpose of clarity, we only visualize the nodes in which DERs are calculated. As can be seen in most nodes, voltage magnitude values calculated by different formulations are close to each other, i.e., the loss of accuracy is lower than 1\%. However, in nodes 23 and 61, the deviation is relatively high and goes over 6\%. In line with the conclusions of the aggregated DER power analysis, voltage magnitudes calculated by the linearised approximation are in general higher, justifying calculated lower export limits, i.e., objective function values.

\begin{figure}
    \centering
    \includegraphics[width=\textwidth]{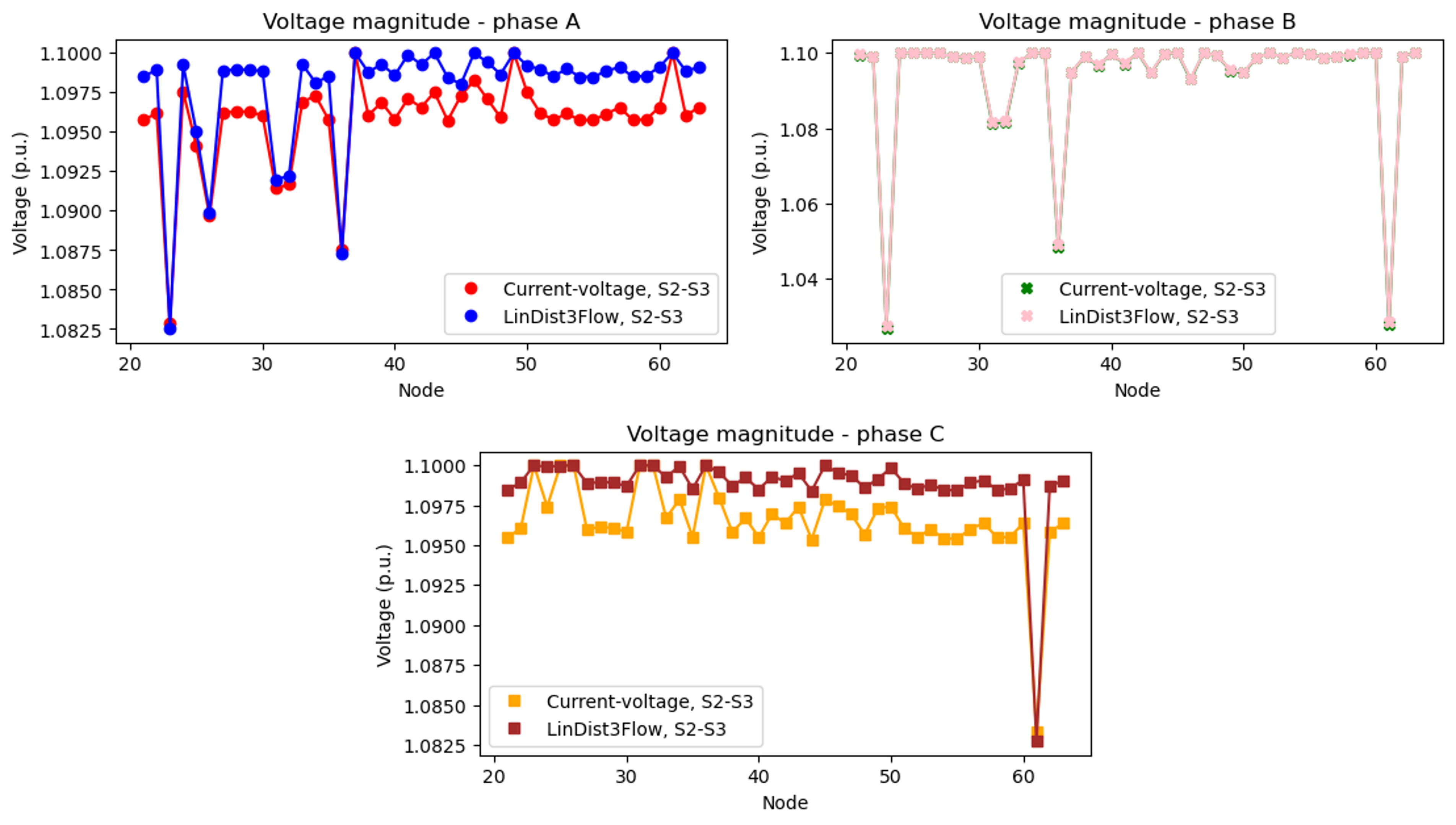}
    \caption{Voltage magnitude in HC calculation, CS 1-1, S2-S3 - Real-world feeder}
    \label{fig:hc_voltage_cs1_s23}
\end{figure}

Differences in voltage magnitude values in the scenario S2-S3 are shown in Figure \ref{fig:hc_voltage_cs1_s23}. In all three phases, the voltage pattern is almost identical, with \textit{LinDist3Flow} values shifted up the y-axis. The accuracy in this scenario is much higher, with the highest deviation between two node voltages of 0.57\%.

\begin{figure}
    \centering
    \includegraphics[width=\textwidth]{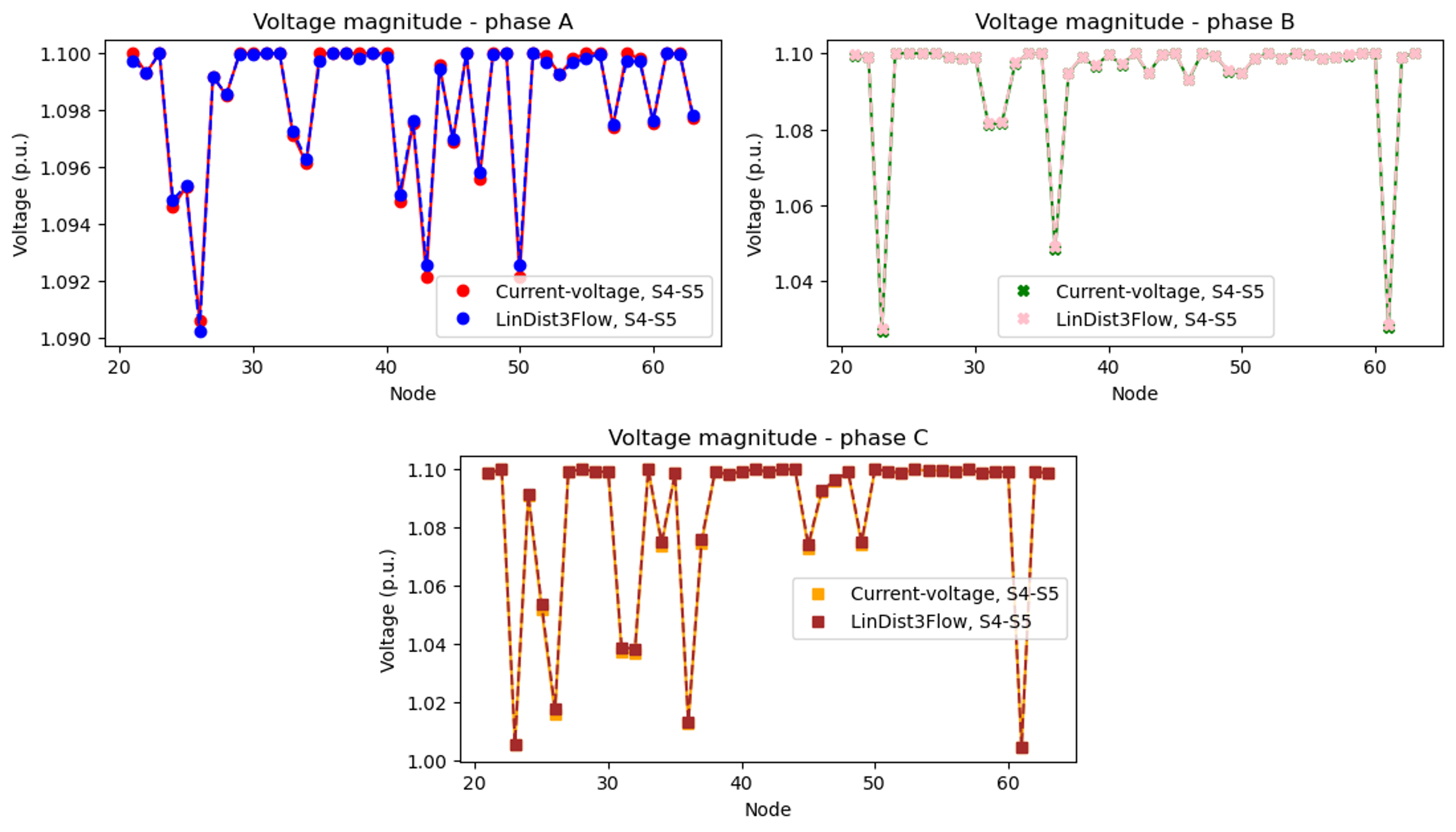}
    \caption{Voltage magnitude in HC calculation, CS 1-1, S4-S5 - Real-world feeder}
    \label{fig:hc_voltage_cs1_s45}
\end{figure}

Figure \ref{fig:hc_voltage_cs1_s45} presents calculated values of phase voltage magnitudes in S4-S5. The maximum deviation between the current-voltage and linearised formulations is lower than 0.1\% in this scenario. Together with the results in S2-S3, this analysis validates the accuracy of linearised formulation in comparison to nonlinear one.

\begin{figure}
    \centering
    \includegraphics[width=\textwidth]{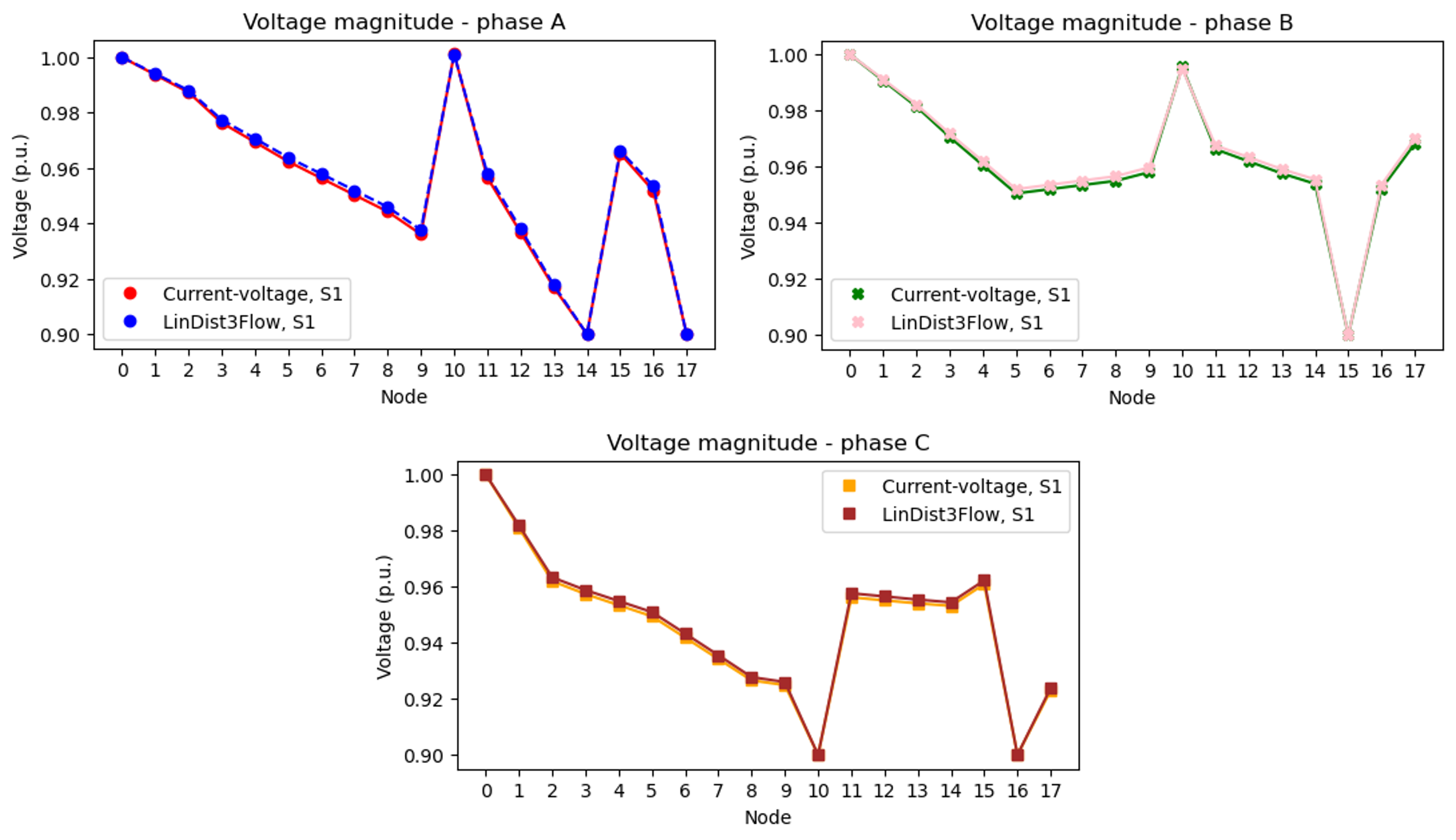}
    \caption{Voltage magnitude in HC calculation, CS 1-2, S1 - CIGRE LV feeder}
    \label{fig:hc_voltage_cs1_2_s1}
\end{figure}

Figure \ref{fig:hc_voltage_cs1_2_s1} shows voltage magnitude values in the case in which binary variables were introduced in the OPF formulation used in the CIGRE LV feeder analysis. Unlike the export calculation, there are no significant differences in phase voltage values in CS 1-2 since the maximum difference is less than 0.1\%. These results confirm the accuracy of the developed MILP formulation. They also suggest that the reason for a higher deviation in CS 1-1 is a different optimality gap set for MINLP and MILP formulations. In the case of different set gaps, the suboptimal solution found can also cause the occurrence of different voltage values and the loss of accuracy.

\begin{figure}
    \centering
    \includegraphics[width=\textwidth]{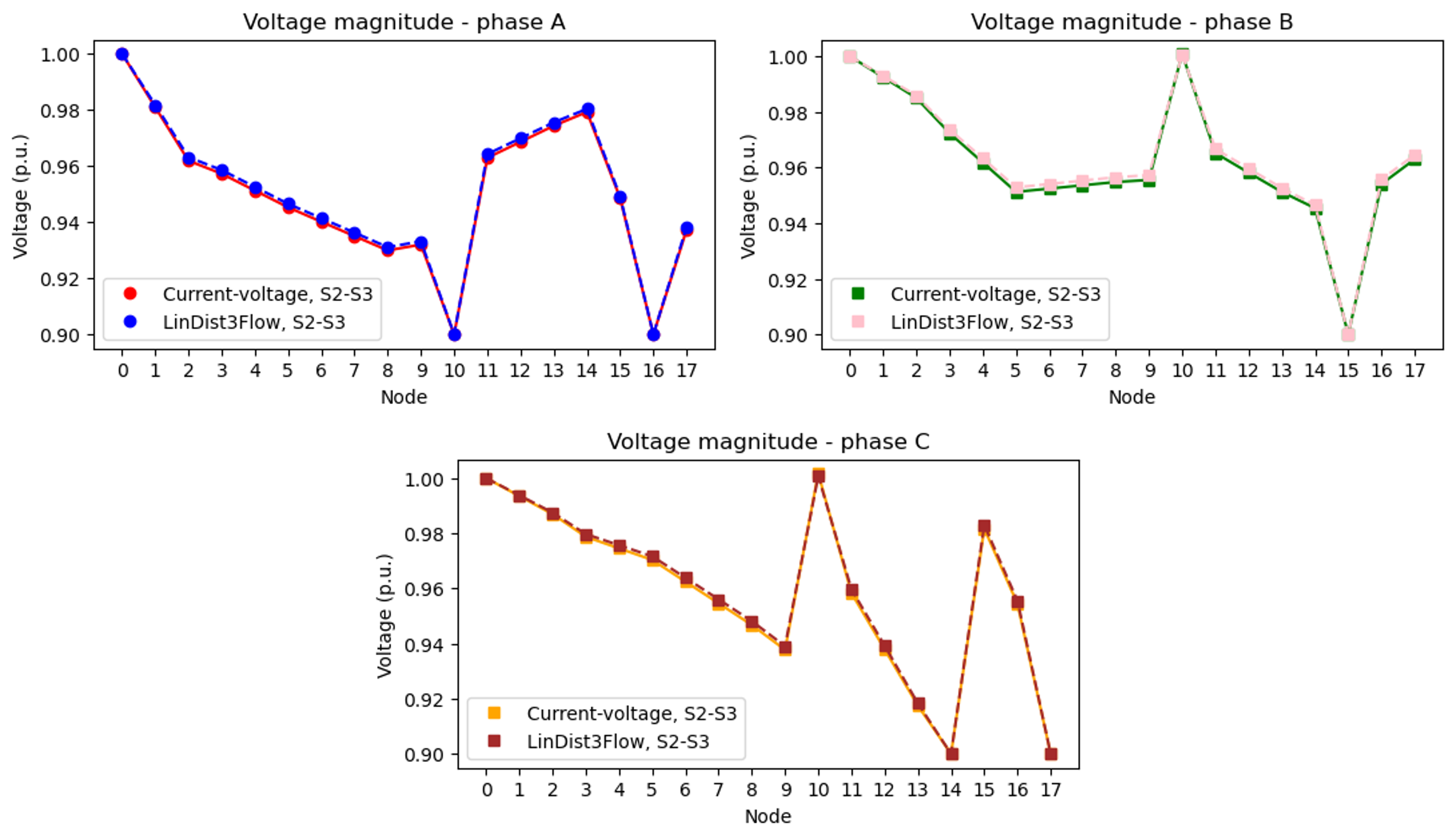}
    \caption{Voltage magnitude in HC calculation, CS 1-2, S2-S3 - CIGRE LV feeder}
    \label{fig:hc_voltage_cs1_2_s23}
\end{figure}

\begin{figure}
    \centering
    \includegraphics[width=\textwidth]{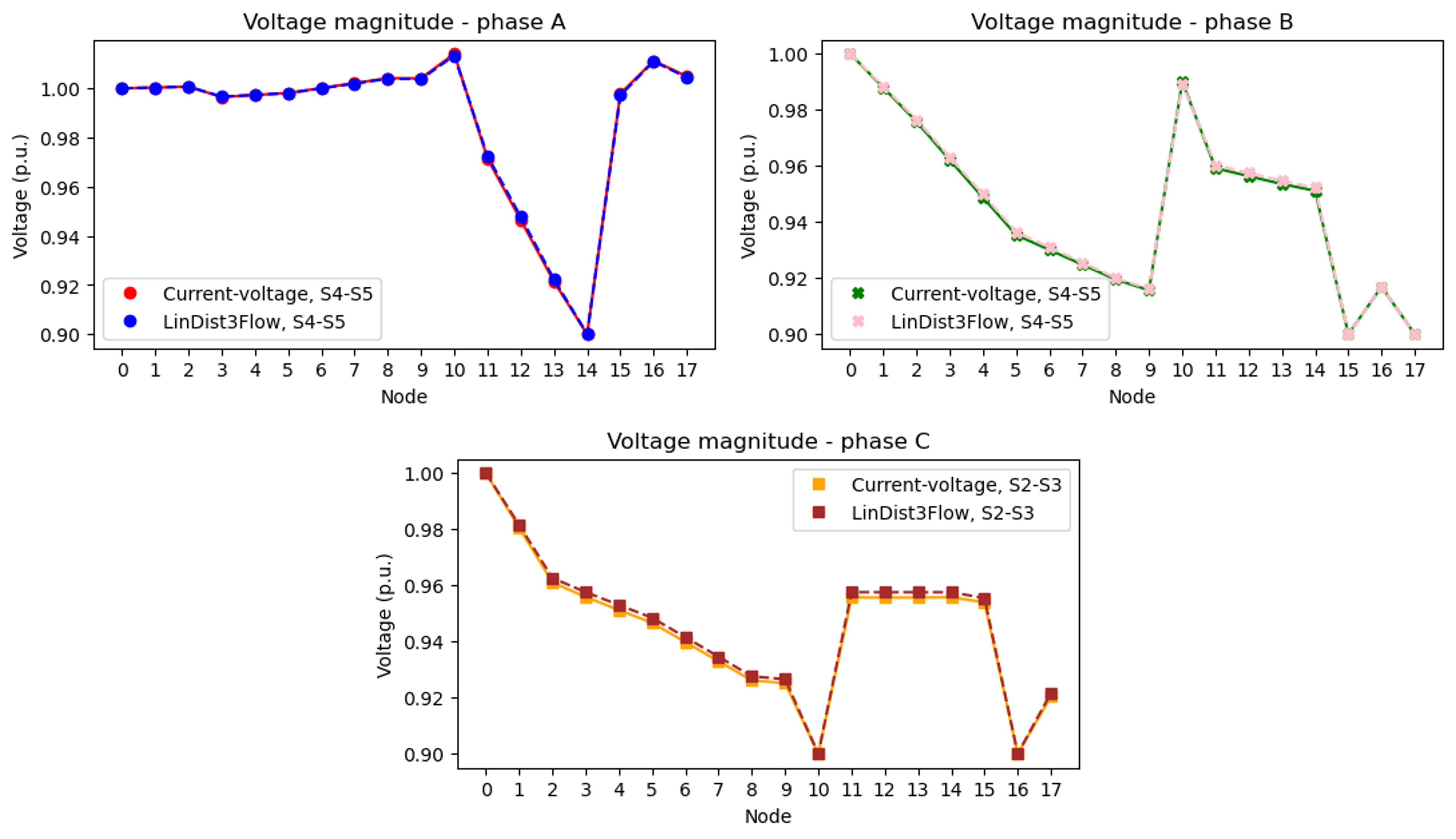}
    \caption{Voltage magnitude in HC calculation, CS 1-2, S4-S5 - CIGRE LV feeder}
    \label{fig:hc_voltage_cs1_2_s45}
\end{figure}

Figures \ref{fig:hc_voltage_cs1_2_s23} and \ref{fig:hc_voltage_cs1_2_s45} show the voltage magnitude values for all three phases calculated in scenarios S2-S3 and S4-S5. The conclusions valid in this scenario remain the same as in the analysis of import limit values, i.e., there are no significant deviations due to the maximum difference of 0.016\%. In general, the voltage values calculated by the linearised formulation are slightly higher than those calculated by the current-voltage formulation. Therefore, the import limits in terms of power are in general higher. That is another validation of the potential inaccuracy the linearisation introduces. The inaccuracy can endanger the safe operation of a network if it is operated near its limits since it can falsely create the space for the adoption of additional loads, which can lead to voltage or current congestion.

\section{Conclusion}\label{sec:conclusion}
With the potential negative impact of uncoordinated DER integration on the planning and operation of distribution networks, it is necessary to calculate DER HC and DOE. These concepts allow the assessment of the maximum installed DER power that does violate the network's technical limitations.
However, both HC and DOE in LV networks are calculated using a three-phase OPF formulation, whose exact model is nonlinear and nonconvex, creating computational complexity, especially in analyses of larger LV networks.
Therefore, many approximation approaches are introduced to decrease the computational time needed to solve the optimisation problem.
In this study, we calculate HC and DOE for single-phase connected DERs. To calculate the optimal values binary variables need to be included in the formulation, uplifting the model to MINLP. Since MINLP problems are nonscalable and intractable, we propose and compare two methods to relax the complexity of the formulation.

In the first approach, we remove binary variables by predetermining the connection phase using different assumptions. In the second approach, we linearise the formulation to MILP and also remove binary variables using the same technique to additionally relax the optimisation problem to LP.
Analyses are performed in the CIGRE LV network and real-world residential LV feeder. The results in both modelled networks show the importance of determining the optimal connection phase since the installed DER power can differ up to 150 kW in the HC calculation and 14 MW in the DOE calculations. However, in a larger real-world LV feeder, computational time becomes problematic since without setting the optimality gap different to zero, optimisation is not finished within the acceptable time, both in cases of the DOE and HC assessment.
The results also show that it is not possible to find a direct correlation between the DER connection phase and objective function value since the results significantly change for case studies in different scenarios.
The importance of the solver is best seen in the case of the MILP formulation, in which gurobi performs much faster than knitro. In all other scenarios, there are no significant differences in computational time when using different solvers.

In the accuracy analysis, we investigate how different test cases, OPF formulations, and scenarios impact per-phase voltage magnitude values. The results show that, in the case of calculating static export limits in a larger network, the results are almost identical in S2-S3 and S4-S5. The exception is S1, in which the optimality gap different from zero needed to be set. Such a setup led to the suboptimal solution, increasing the difference between voltage magnitude values.
In the case of import limits calculated in the CIGRE LV network, there are no significant differences since the maximum error does not exceed 0.16\%. The importance of using an adequate solver is also investigated in the paper. The results show that not all solvers are suitable for a defined problem since only knitro was able to solve the MINLP problem. Knitro also solves the NLP formulation faster than ipopt, while gurobi is not suitable for solving nonlinear and non-convex problems. Ipopt is also the slowest in solving linear problems, while knitro is the fastest, except in the MILP formulation, being solved fastest by gurobi.

\section*{Acknowledgment}
This work has been in part supported by the the WeForming project. The WeForming project has received funding from the European Union’s Horizon Europe Programme under the Grant Agreement No. 101123556.

\bibliography{literature}

\begin{thebibliography}{10}
\providecommand{\url}[1]{#1}
\csname url@samestyle\endcsname
\providecommand{\newblock}{\relax}
\providecommand{\bibinfo}[2]{#2}
\providecommand{\BIBentrySTDinterwordspacing}{\spaceskip=0pt\relax}
\providecommand{\BIBentryALTinterwordstretchfactor}{4}
\providecommand{\BIBentryALTinterwordspacing}{\spaceskip=\fontdimen2\font plus
\BIBentryALTinterwordstretchfactor\fontdimen3\font minus \fontdimen4\font\relax}
\providecommand{\BIBforeignlanguage}[2]{{%
\expandafter\ifx\csname l@#1\endcsname\relax
\typeout{** WARNING: IEEEtran.bst: No hyphenation pattern has been}%
\typeout{** loaded for the language `#1'. Using the pattern for}%
\typeout{** the default language instead.}%
\else
\language=\csname l@#1\endcsname
\fi
#2}}
\providecommand{\BIBdecl}{\relax}
\BIBdecl

\bibitem{IRENA2024}
\BIBentryALTinterwordspacing
IRENA, ``{Renewable capacity statistics 2024},'' International Renewable Energy Agency, Abu Dhabi, Tech. Rep., 2024. [Online]. Available: \url{https://www.irena.org/Publications/2024/Mar/Renewable-capacity-statistics-2024}
\BIBentrySTDinterwordspacing

\bibitem{IEA2022}
\BIBentryALTinterwordspacing
IEA, ``{Approximately 100 million households rely on rooftop solar PV by 2030},'' 2022. [Online]. Available: \url{https://www.iea.org/reports/approximately-100-million-households-rely-on-rooftop-solar-pv-by-2030}
\BIBentrySTDinterwordspacing

\bibitem{IEA2023}
\BIBentryALTinterwordspacing
------, ``{World Energy Outlook 2023},'' International Energy Agency, Paris, Tech. Rep., 2023. [Online]. Available: \url{https://www.iea.org/reports/world-energy-outlook-2023}
\BIBentrySTDinterwordspacing

\bibitem{IEA2023EV}
\BIBentryALTinterwordspacing
------, ``{Global EV Outlook 2023},'' International Energy Agency, Paris, Tech. Rep., 2023. [Online]. Available: \url{https://www.iea.org/reports/global-ev-outlook-2023}
\BIBentrySTDinterwordspacing

\bibitem{ANTIC2022100926}
T.~Antić and T.~Capuder, ``{Utilization of physical devices for the improvement of power quality indicators during the COVID-19 pandemic and uncoordinated integration of low carbon units},'' \emph{Sustainable Energy, Grids and Networks}, vol.~32, p. 100926, 2022.

\bibitem{grid_code}
\BIBentryALTinterwordspacing
``{Croatian distribution system grid code (Mrežna pravila distribucijskog sustava)},'' 2018. [Online]. Available: \url{https://narodne-novine.nn.hr/clanci/sluzbeni/2018\_08\_74\_1539.html}
\BIBentrySTDinterwordspacing

\bibitem{10415382}
H.~H.~H. Mousa, K.~Mahmoud, and M.~Lehtonen, ``{A Comprehensive Review on Recent Developments of Hosting Capacity Estimation and Optimization for Active Distribution Networks},'' \emph{IEEE Access}, vol.~12, pp. 18\,545--18\,593, 2024.

\bibitem{YAO2022119681}
H.~Yao, W.~Qin, X.~Jing, Z.~Zhu, K.~Wang, X.~Han, and P.~Wang, ``Possibilistic evaluation of photovoltaic hosting capacity on distribution networks under uncertain environment,'' \emph{Applied Energy}, vol. 324, p. 119681, 2022.

\bibitem{EDMUNDS2021117093}
C.~Edmunds, S.~Galloway, J.~Dixon, W.~Bukhsh, and I.~Elders, ``Hosting capacity assessment of heat pumps and optimised electric vehicle charging on low voltage networks,'' \emph{Applied Energy}, vol. 298, p. 117093, 2021.

\bibitem{FACHRIZAL2021100445}
R.~Fachrizal, U.~H. Ramadhani, J.~Munkhammar, and J.~Widén, ``Combined pv–ev hosting capacity assessment for a residential lv distribution grid with smart ev charging and pv curtailment,'' \emph{Sustainable Energy, Grids and Networks}, vol.~26, p. 100445, 2021.

\bibitem{en14010117}
T.~Antić, T.~Capuder, and M.~Bolfek, ``{A Comprehensive Analysis of the Voltage Unbalance Factor in PV and EV Rich Non-Synthetic Low Voltage Distribution Networks},'' \emph{Energies}, vol.~14, no.~1, 2021.

\bibitem{10102721}
Y.~Hou, J.~Zhu, M.~Z. Liu, W.~J. Nacmanson, and L.~F. Ochoa, ``{EV Hosting Capacity and Voltage Unbalance: An Australian Case Study},'' in \emph{2023 IEEE PES Grid Edge Technologies Conference \& Exposition (Grid Edge)}, 2023, pp. 1--5.

\bibitem{TOGHRANEGAR2022104243}
S.~Toghranegar, A.~Rabiee, and A.~Soroudi, ``{Enhancing the unbalanced distribution network’s hosting capacity for DERs via optimal load re-phasing},'' \emph{Sustainable Cities and Society}, vol.~87, p. 104243, 2022.

\bibitem{10257450}
T.~Antić, A.~Nouri, A.~Keane, and T.~Capuder, ``{Solving Scalability Issues in Calculating PV Hosting Capacity in Low Voltage Distribution Networks},'' in \emph{2023 International Conference on Smart Energy Systems and Technologies (SEST)}, 2023, pp. 1--6.

\bibitem{GETH2020106558}
F.~Geth, S.~Claeys, and G.~Deconinck, ``{Nonconvex lifted unbalanced branch flow model: Derivation, implementation and experiments},'' \emph{Electric Power Systems Research}, vol. 189, p. 106558, 2020.

\bibitem{7038399}
L.~Gan and S.~H. Low, ``Convex relaxations and linear approximation for optimal power flow in multiphase radial networks,'' in \emph{2014 Power Systems Computation Conference}, 2014, pp. 1--9.

\bibitem{7741261}
D.~B. Arnold, M.~Sankur, R.~Dobbe, K.~Brady, D.~S. Callaway, and A.~Von~Meier, ``Optimal dispatch of reactive power for voltage regulation and balancing in unbalanced distribution systems,'' in \emph{2016 IEEE Power and Energy Society General Meeting (PESGM)}, 2016, pp. 1--5.

\bibitem{7244261}
B.~A. Robbins and A.~D. Domínguez-García, ``{Optimal Reactive Power Dispatch for Voltage Regulation in Unbalanced Distribution Systems},'' \emph{IEEE Transactions on Power Systems}, vol.~31, no.~4, pp. 2903--2913, 2016.

\bibitem{9639999}
S.~Claeys, F.~Geth, M.~Sankur, and G.~Deconinck, ``{No-Load Linearization of the Lifted Multi-Phase Branch Flow Model: Equivalence and Case Studies},'' in \emph{2021 IEEE PES Innovative Smart Grid Technologies Europe (ISGT Europe)}, 2021, pp. 1--5.

\bibitem{VANIN2020106699}
M.~Vanin, H.~Ergun, R.~D’hulst, and D.~{Van Hertem}, ``{Comparison of Linear and Conic Power Flow Formulations for Unbalanced Low Voltage Network Optimization},'' \emph{Electric Power Systems Research}, vol. 189, p. 106699, 2020.

\bibitem{5491276}
R.~D. Zimmerman, C.~E. Murillo-Sánchez, and R.~J. Thomas, ``{MATPOWER: Steady-State Operations, Planning, and Analysis Tools for Power Systems Research and Education},'' \emph{IEEE Transactions on Power Systems}, vol.~26, no.~1, pp. 12--19, 2011.

\bibitem{openDSS}
\BIBentryALTinterwordspacing
{Electric Power Research Institute (EPRI)}, ``{OpenDSS, Open Distribution System Simulator}.'' [Online]. Available: \url{https://sourceforge.net/projects/electricdss/files/}
\BIBentrySTDinterwordspacing

\bibitem{8344496}
L.~Thurner, A.~Scheidler, F.~Schäfer, J.-H. Menke, J.~Dollichon, F.~Meier, S.~Meinecke, and M.~Braun, ``{Pandapower—An Open-Source Python Tool for Convenient Modeling, Analysis, and Optimization of Electric Power Systems},'' \emph{IEEE Transactions on Power Systems}, vol.~33, no.~6, pp. 6510--6521, 2018.

\bibitem{9282125}
V.~Rigoni and A.~Keane, ``{Open-DSOPF: an open-source optimal power flow formulation integrated with OpenDSS},'' in \emph{2020 IEEE Power \& Energy Society General Meeting (PESGM)}, 2020, pp. 1--5.

\bibitem{10252870}
T.~Antić, A.~Keane, and T.~Capuder, ``{Pp OPF -Pandapower Implementation of Three-phase Optimal Power Flow Model},'' in \emph{2023 IEEE Power \& Energy Society General Meeting (PESGM)}, 2023, pp. 1--5.

\bibitem{FOBES2020106664}
D.~M. Fobes, S.~Claeys, F.~Geth, and C.~Coffrin, ``{PowerModelsDistribution.jl: An open-source framework for exploring distribution power flow formulations},'' \emph{Electric Power Systems Research}, vol. 189, p. 106664, 2020.

\bibitem{ppOPF}
\BIBentryALTinterwordspacing
``{ppOPF}.'' [Online]. Available: \url{https://github.com/tomislavantic/ppOPF}
\BIBentrySTDinterwordspacing

\bibitem{9715663}
T.~Milford and O.~Krause, ``{Managing DER in Distribution Networks Using State Estimation \& Dynamic Operating Envelopes},'' in \emph{2021 IEEE PES Innovative Smart Grid Technologies - Asia (ISGT Asia)}, 2021, pp. 1--5.

\bibitem{9816082}
M.~Z. Liu, L.~F. Ochoa, P.~K.~C. Wong, and J.~Theunissen, ``{Using OPF-Based Operating Envelopes to Facilitate Residential DER Services},'' \emph{IEEE Transactions on Smart Grid}, vol.~13, no.~6, pp. 4494--4504, 2022.

\bibitem{10202795}
T.~Antić, F.~Geth, and T.~Capuder, ``{The Importance of Technical Distribution Network Limits in Dynamic Operating Envelopes},'' in \emph{2023 IEEE Belgrade PowerTech}, 2023, pp. 1--6.

\bibitem{9133699}
F.~Geth, S.~Claeys, and G.~Deconinck, ``{Current-Voltage Formulation of the Unbalanced Optimal Power Flow Problem},'' in \emph{2020 8th Workshop on Modeling and Simulation of Cyber-Physical Energy Systems}, 2020, pp. 1--6.

\bibitem{en50160}
``{EN 50160 Voltage characteristic of electricity supplied by public electricity networks},'' p.~34, 2010.

\bibitem{iec61000_2_2}
``{Electromagnetic compatibility (EMC) - Part 2-2: Environment - Compatibility levels for low-frequency conducted disturbances and signalling in public low-voltage power supply systems},'' p.~57, 2002.

\bibitem{CIGRE_LV}
K.~Strunz, E.~Abbasi, R.~Fletcher, N.~Hatziargyriou, R.~Iravani, and G.~Joos, \emph{{TF C6.04.02: TB 575 - Benchmark Systems for Network Integration of Renewable and Distributed Energy Resources}}.\hskip 1em plus 0.5em minus 0.4em\relax CIGRE, April 2014.

\bibitem{Hart2011}
W.~E. Hart, J.-P. Watson, and D.~L. Woodruff, ``{Pyomo: modeling and solving mathematical programs in Python},'' \emph{Mathematical Programming Computation}, vol.~3, no.~3, pp. 219--260, Sep 2011.

\end{thebibliography}
\bibliographystyle{IEEEtran}

\end{document}